\documentclass[10pt,journal]{IEEEtran}

\usepackage{epsfig,latexsym}
\usepackage{float}
\usepackage{indentfirst}
\usepackage{amsmath}
\usepackage{bm}
\usepackage{amssymb}
\usepackage{times}
\usepackage{enumitem}
\usepackage{cite}
\usepackage{lastpage}
\usepackage{color}
\usepackage{amsthm}
\usepackage{array}
\usepackage{algorithm}
\usepackage{algorithmic}
\usepackage{multirow}
\usepackage{empheq}
\usepackage{graphicx}
\usepackage[caption=false,font=footnotesize]{subfig}
\usepackage{booktabs}

\newtheorem{lemma}{Lemma}

\newtheorem{theorem}{Theorem}

\begin{document}

\title{Dynamic Metasurface-Enabled Generalized Pinching-Antenna Systems}
\author{Shan Shan, Chongjun Ouyang, Yufei Zhao, Zhiguo Ding, and Yong Liang Guan%
\thanks{Shan Shan, Yufei Zhao, Zhiguo Ding, and Yong Liang Guan are with the School of Electrical and Electronic Engineering, Nanyang Technological University, Singapore 639798 (e-mail: {\{shan.shan, yufei.zhao, zhiguo.ding, eylguan\}}@ntu.edu.sg). Chongjun Ouyang is with the School of Electronic Engineering and Computer Science, Queen Mary University of London, London E1 4NS, U.K. (email: c.ouyang@qmul.ac.uk).}
}
\IEEEaftertitletext{\vspace{-3em}}
\maketitle

\begin{abstract}
This article investigates dynamic metasurfaces as an electronic realization of generalized pinching antennas (GPA). 
By jointly optimizing the locations and radiation coefficients of activated elements, the dynamic metasurface-enabled GPA (DM-GPA) architecture exploits additional design flexibility to enhance spatial reconfigurability beyond that attainable through antenna placement in the conventional pinching-antenna system (PASS).
Based on this architecture, we study weighted sum rate (WSR) maximization for multiuser downlink transmission.
1) For two-user transmission, the sum-rate improvement of DM-GPA over conventional PASS is quantified through a lower bound that reveals when the benefit of channel orthogonalization achieved by adjusting the radiation coefficients outweighs the channel power loss caused by the Lorentzian constraint. Furthermore, when conventional PASS yields a rank-deficient channel, DM-GPA is shown to achieve spatial multiplexing and transmit diversity, with a strict sum-rate advantage at sufficiently high signal-to-noise ratios.
2) For general multiuser transmission, a weighted minimum mean-square error (WMMSE)-based alternating optimization algorithm is developed for joint pinching beamforming and digital precoder design. A sequential element-wise search is developed to optimize the locations of the activated elements, with a closed-form solution for the radiation coefficient at each candidate location.
Numerical results are provided to demonstrate that: (i) DM-GPA achieves higher WSR than other PASS architectures and fixed-array benchmarks, (ii) increasing the number of activated elements further improves the WSR of DM-GPA, and (iii) radiation coefficient adjustment is particularly beneficial when candidate activation locations are limited.
\end{abstract}


\section{Introduction}
\label{sec:introduction}
Flexible-antenna technologies have attracted growing interest for their ability to reconfigure wireless channels through antenna positioning or electromagnetic control~\cite{10910066}. Representative approaches include fluid and movable antennas, which improve channel conditions through local position adjustments~\cite{9264694,10286328}, and reconfigurable intelligent surfaces (RISs), which modify reflected propagation paths~\cite{9424177}. However, position adjustments in conventional fluid- and movable-antenna systems are typically confined to compact regions, which limits their ability to shorten propagation distances to distant users or circumvent large obstacles. Passive RIS-assisted links remain subject to attenuation along both the transmitter-to-RIS and RIS-to-user paths. These limitations motivate antenna architectures that can reconfigure radiation locations over an extended service region and bring wireless access closer to randomly distributed users.

The pinching-antenna system (PASS) enables such large-scale spatial reconfiguration through adjustable radiation points along dielectric waveguides~\cite{10945421}. Small dielectric particles, termed \emph{pinching antennas (PAs)}, are attached at selected positions along a waveguide to couple guided radio-frequency (RF) signals into free space~\cite{fukuda2022pinching}. Since dielectric waveguides support low-loss transmission over an extended service region, PAs can be deployed close to users to establish strong line-of-sight (LoS) links and reduce free-space path loss. PASS therefore offers large-scale antenna reconfiguration, last-meter wireless access, and scalable deployment with low hardware complexity~\cite{11169486}.


Motivated by these advantages, extensive research has been conducted on PASS-enabled wireless communications~\cite{11169486}. In particular, antenna placement has been widely investigated to reduce path loss and improve transmission rates. The authors in~\cite{10896748} studied downlink rate maximization and proposed a two-stage PA placement algorithm based on path-loss minimization. To reduce the complexity of antenna placement optimization, the authors in~\cite{11016750} derived closed-form placement solutions for orthogonal and non-orthogonal multiple access. For multiuser transmission, the authors in~\cite{11048566} proposed a two-timescale scheme that jointly optimizes long-term PA placement and short-term digital precoding.
Beyond antenna placement, studies have also incorporated radiation power distribution into PASS modeling. Specifically, the authors in~\cite{wang2025modeling} developed physics-based equal-power and proportional-power models and jointly optimized transmit and pinching beamforming under continuous and discrete location constraints. The work in~\cite{11263923} further controls radiation power by adjusting the coupling spacing between each PA and the waveguide. 

Realizing such control over radiation locations and power also depends on the physical antenna implementation. The generalized pinching-antenna (GPA) framework extends PASS beyond its original dielectric-waveguide implementation to alternative transmission media and signal-distribution architectures~\cite{xu2026generalizedtutorial}. Representative examples include leaky-coaxial-cable GPA (LCX-GPA), which uses controllable radiating slots along a coaxial cable~\cite{wang2025lcxgpa}, and radio-stripe GPA (RS-GPA), which employs active antenna processing units (APUs) distributed along a shared cable~\cite{xu2026rsgpa}. The segmented waveguide-enabled pinching-antenna system (SWAN) uses multiple short dielectric waveguide segments, each with its own feed point and a wired connection to the base station~\cite{ouyang2026swan}. Electronically reconfigurable pinching antennas have also been proposed to control radiation from dielectric waveguides through varactor-loaded elements~\cite{plata2026epinching}. These implementations illustrate the adaptability of the GPA framework to different deployment scenarios and hardware requirements.

\subsection{Motivation and Contributions}
Practical PASS deployment requires both a suitable antenna architecture and an effective mechanism for adjusting antenna locations. Several existing studies assume continuously adjustable PA locations and optimize them for phase alignment~\cite{10896748,11016750}. Implementing such designs with movable PAs requires precise mechanical repositioning along the waveguide. A practical alternative is to activate pre-installed elements at discrete candidate locations~\cite{wang2025modeling}. Elements near users can then be activated to reduce path loss without mechanical repositioning. However, the finite candidate set restricts the available phase adjustments, and locations with favorable path loss do not necessarily provide the phase relationships required for coherent combining and inter-user interference suppression~\cite{wang2025modeling}. This limitation motivates additional phase control at the selected locations.

{Dynamic metasurfaces provide this additional control through electronically tunable waveguide-fed meta-atoms~\cite{smith2017waveguide,jabbar2024dma}.} Existing studies have jointly designed the radiation coefficients and digital processing for uplink reception and multiuser downlink transmission at prescribed element locations~\cite{shlezinger2019dma,kimaryo2023downlink}. Motivated by this capability, we propose a dynamic-metasurface-enabled GPA (DM-GPA) architecture in which a subset of pre-installed meta-atoms is activated along each waveguide. Each activated element has an adjustable radiation coefficient whose amplitude and phase are coupled by the Lorentzian constraint. Adjusting the radiation coefficients of the activated elements provides an additional degree of spatial reconfigurability. To exploit the flexibility in both location selection and radiation control, we investigate joint pinching beamforming and digital precoder design for weighted sum rate (WSR) maximization. The main contributions are summarized as follows:

\begin{itemize}
    \item We propose a DM-GPA-enabled multiuser downlink framework that realizes pinching beamforming through selective activation of pre-installed elements and adjustment of their radiation coefficients. Based on this system model, we formulate a WSR maximization problem by jointly optimizing pinching beamforming and the digital precoder.

    \item We establish an analytical comparison with conventional PASS at identical element locations. For two-user transmission, we derive a closed-form orthogonalization design and a lower bound on the resulting sum-rate improvement. The bound quantifies the tradeoff between interference suppression and channel power loss under the Lorentzian constraint. For rank-deficient conventional PASS channels, we further show that the orthogonal design supports an additional spatial stream and achieves a strictly higher sum rate at sufficiently high transmit powers. For transmit diversity, we derive the maximum receive signal-to-noise ratio (SNR) in closed form and quantify the gain from coherent combining within each waveguide.

    \item For general multiuser transmission, we develop an alternating optimization (AO) algorithm based on an equivalent weighted minimum mean-square error (WMMSE) formulation. For digital precoding, we absorb the transmit power constraint into the effective noise terms through power normalization and solve the resulting unconstrained quadratic subproblem in closed form. For pinching beamforming, we develop a sequential element-wise search to optimize the locations of the activated elements, with the radiation coefficient optimized in closed form at each candidate location.

    \item We present numerical results to validate the performance advantage of DM-GPA and the effectiveness of the proposed algorithm. The results demonstrate that (i) DM-GPA achieves higher WSR than the benchmark schemes and maintains this advantage as the service region expands, (ii) increasing the number of activated elements per waveguide further improves WSR, and (iii) radiation coefficient optimization yields larger WSR improvements when antenna placement is restricted to a limited set of candidate locations.
\end{itemize}

\subsection{Organization and Notations}

The remainder of this paper is organized as follows. Section \ref{sec:system_model} introduces the DM-GPA system model and formulates the WSR maximization problem. Section \ref{sec:two_user_analysis} analyzes the performance advantages of DM-GPA over conventional PASS for two-user transmission. Section \ref{sec:joint_optimization} presents the joint pinching beamforming and digital precoder design for general multiuser transmission. Numerical results are presented in Section \ref{sec:numerical_results}. Finally, Section \ref{sec:conclusion} concludes this paper.

\emph{Notations}: Bold lowercase and uppercase letters denote vectors and matrices, respectively. The operators $(\cdot)^*$, $(\cdot)^{\rm T}$, and $(\cdot)^{\rm H}$ denote complex conjugation, transpose, and conjugate transpose, respectively. The symbols $\|\cdot\|_2$ and $\|\cdot\|_{\rm F}$ denote the Euclidean and Frobenius norms, respectively. The operators $\mathbb E[\cdot]$, $\operatorname{Tr}(\cdot)$, and $\operatorname{Re}\{\cdot\}$ denote expectation, trace, and real part, respectively. The operators $\operatorname{diag}(\cdot)$ and $\operatorname{blkdiag}(\cdot)$ construct diagonal and block-diagonal matrices, respectively. The notation $\mathbf A\succeq\mathbf B$ means that $\mathbf A-\mathbf B$ is positive semidefinite, while $\mathbf v\succ\mathbf0$ denotes entrywise positivity. The notation $\mathcal{CN}(0,\sigma^2)$ denotes a zero-mean circularly symmetric complex Gaussian distribution with variance $\sigma^2$. For real $x$, $[x]_+=\max\{x,0\}$.

\section{System Model}
\label{sec:system_model}

\begin{figure}[!t]
\centering
\includegraphics[width=0.48\textwidth]{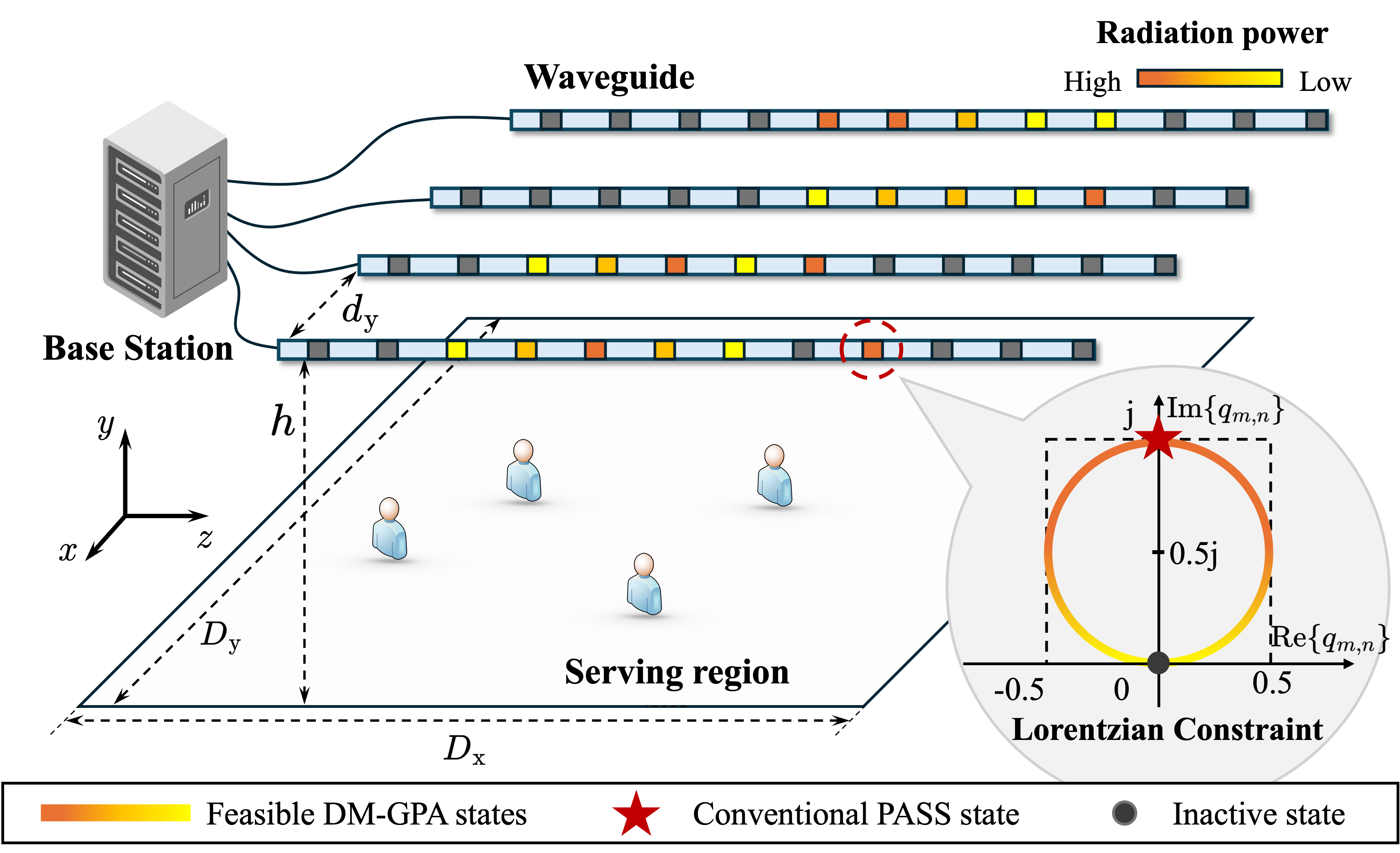} 
\caption{Illustration of the proposed DM-GPA architecture. The conventional PASS baseline fixes the radiation coefficient of every activated element at \(q_{m,n}=\mathrm{j}\), whereas \(q_{m,n}=0\) switches off an element.}
\label{fig:system_model}
\end{figure}

In this section, we consider DM-GPA-enabled multiuser downlink transmission. As illustrated in Fig.~\ref{fig:system_model}, the base station (BS) is equipped with $M$ parallel waveguides to serve $K$ single-antenna users. Let $\mathcal{K}\triangleq\{1,\ldots,K\}$ denote the user set. The users are independently and uniformly distributed over a rectangular ground region of size $D_x\times D_y$ m$^2$. The position of the $k$th user is denoted by $\mathbf u_k=[x_k,y_k,0]^{\rm T}$, $k\in\mathcal{K}$, where $x_k\in[0,D_x]$ and $y_k\in[-D_y/2,D_y/2]$. The user locations are assumed to be known at the BS.
\vspace{-5pt}
\subsection{DM-GPA Architecture}
\label{subsec:architecture}

The proposed DM-GPA architecture uses $M$ parallel waveguides, each fed by a dedicated RF chain carrying a digitally precoded signal from the BS. Each waveguide contains $Q$ pre-installed, electronically tunable meta-atoms that serve as radiating elements. The activated meta-atoms couple part of the guided signal into free space, and their radiation depends on both in-waveguide propagation and the tunable radiation coefficients.

We first specify the waveguide layout and the locations available for element activation. The waveguides are deployed at a fixed height $h$ above the serving region and extend along the $x$-axis from $0$ to $D_x$. Let $\mathcal{M}\triangleq\{1,\ldots,M\}$ denote the waveguide index set. For $M>1$, the waveguides are uniformly spaced along the $y$-axis with an interval of $d=D_y/(M-1)$, and the $y$-coordinate of the $m$th waveguide is $y_m=-D_y/2+(m-1)d$. The feed point of this waveguide is located at $\mathbf r_{m,0}=[0,y_m,h]^{\rm T}$.

On each waveguide, the $Q$ pre-installed meta-atoms occupy the discrete candidate locations in a common set $\mathcal{S}_x\subset[0,D_x]$. The radiation locations are selected by activating $N\leq Q$ elements and keeping the remaining elements inactive. The activated elements are indexed by $\mathcal{N}\triangleq\{1,\ldots,N\}$. The position of the $n$th activated element on the $m$th waveguide is denoted by $\mathbf r_{m,n}=[x_{m,n},y_m,h]^{\rm T}$, $m\in\mathcal{M}$ and $n\in\mathcal{N}$. Here, $x_{m,n}\in\mathcal{S}_x$ represents the distance along the $x$-axis from the feed point to this element.
We collect these selected locations in $\mathbf X=[\mathbf x_1,\ldots,\mathbf x_M]^{\rm T}\in\mathbb R^{M\times N}$, where $\mathbf x_m=[x_{m,1},\ldots,x_{m,N}]^{\rm T}$ contains the locations of the activated elements on the $m$th waveguide. The feasible set $\mathcal{X}$ consists of matrices whose rows contain $N$ distinct locations from $\mathcal{S}_x$.

At the selected locations, the radiation amplitude and phase are controlled electronically. Each meta-atom is modeled as a tunable resonant circuit whose electrical properties are adjusted by a dedicated controller~\cite{smith2017waveguide,shlezinger2019dma}. For the considered narrowband transmission, the frequency response of each meta-atom is assumed to be flat over the signal bandwidth~\cite{shlezinger2019dma,kimaryo2023downlink}. Under this model, the complex radiation coefficient $q_{m,n}$ of the $n$th activated meta-atom on the $m$th waveguide satisfies the Lorentzian constraint illustrated in Fig.~\ref{fig:system_model} and is given by
\begin{equation}
 q_{m,n}=\frac{\mathrm{j}+\exp(\mathrm{j}\theta_{m,n})}{2}\in\mathcal{Q}_{\rm L},
\label{eq:lorentzian_theta}
\end{equation}
where $\theta_{m,n}\in[0,2\pi)$ is independently tunable for each element and $\mathcal{Q}_{\rm L}$ denotes the feasible set of radiation coefficients. The radiation amplitude and phase are coupled by \eqref{eq:lorentzian_theta}, with $0\leq|q_{m,n}|\leq1$. In particular, $q_{m,n}=0$ switches off radiation. At the same selected locations, fixing $q_{m,n}=\mathrm{j}$ for every activated element recovers the conventional PASS baseline, whose radiation coefficients have unit magnitude and a common fixed phase of $\pi/2$.
To parameterize the radiation coefficients for subsequent optimization, we introduce $z_{m,n}=\exp(\mathrm{j}\theta_{m,n})$, which satisfies $|z_{m,n}|=1$ and gives $q_{m,n}=(\mathrm{j}+z_{m,n})/2$. These variables are collected in $\mathbf Z=[z_{m,n}]\in\mathbb C^{M\times N}$.

\vspace{-8pt}
\subsection{Channel Model}
\label{subsec:dma_channel}

Within the $m$th waveguide, the signal from the feed point $\mathbf r_{m,0}$ to the $n$th activated element $\mathbf r_{m,n}$ experiences attenuation and a phase shift determined by $x_{m,n}$. The corresponding in-waveguide propagation coefficient can be given by
\begin{equation}
 p_{m,n}=\frac{1}{\sqrt N}10^{-\epsilon x_{m,n}/20}
 \exp(-\mathrm{j}\beta_{\rm g}x_{m,n}),
\label{eq:guided_response}
\end{equation}
where $\epsilon$ is the in-waveguide attenuation in dB/m and $\beta_{\rm g}=2\pi/\lambda_{\rm g}$ denotes the guided wavenumber. The guided wavelength is $\lambda_{\rm g}=\lambda/n_{\rm eff}$, where $\lambda$ is the free-space wavelength and $n_{\rm eff}$ is the waveguide's effective refractive index. The factor $1/\sqrt N$ represents equal feed-power scaling across the $N$ activated elements. These propagation coefficients are collected in the in-waveguide channel vector $\mathbf p_m(\mathbf x_m)=[p_{m,1},\ldots,p_{m,N}]^{\rm T}\in\mathbb C^N$ for the $m$th waveguide. Since each RF chain feeds one waveguide, the in-waveguide channel matrix has a block-diagonal structure and is given by
\begin{equation}
 \mathbf P(\mathbf X)
 =\operatorname{blkdiag}\bigl(\mathbf p_1(\mathbf x_1),\ldots,\mathbf p_M(\mathbf x_M)\bigr)
 \in\mathbb C^{MN\times M}.
\label{eq:guided_channel_matrix}
\end{equation}

As in conventional PASS, the radiation locations in DM-GPA can be selected close to users to establish short LoS links. We therefore adopt a spherical-wave free-space channel model~\cite{10945421,wang2025modeling} to analyze the effects of location selection and radiation coefficient adjustment on multiuser transmission. Let $r_{k,m,n}=\|\mathbf u_k-\mathbf r_{m,n}\|_2$ denote the distance from the $n$th activated element on the $m$th waveguide to the $k$th user. The corresponding channel coefficient is given as follows:
\begin{equation}
 h_{k,m,n}=\frac{\sqrt{\eta}}{r_{k,m,n}}
 \exp(-\mathrm{j} k_0r_{k,m,n}),
\label{eq:free_space_channel}
\end{equation}
where $k_0=2\pi/\lambda$ is the free-space wavenumber and $\eta=(\lambda/4\pi)^2$ is the reference free-space power gain. The channel from the $m$th waveguide to the $k$th user is represented by $\mathbf h_{k,m}^{\rm H}(\mathbf x_m)=[h_{k,m,1},\ldots,h_{k,m,N}]$. The complete user channel vector and the free-space channel matrix are then obtained as follows:
\begin{align}
 \mathbf h_k^{\rm H}(\mathbf X)
 &=[\mathbf h_{k,1}^{\rm H}(\mathbf x_1),\ldots,\mathbf h_{k,M}^{\rm H}(\mathbf x_M)]\in\mathbb C^{1\times MN},
\label{eq:free_space_channel_vector}\\
 \mathbf H(\mathbf X)
 &=[\mathbf h_1(\mathbf X),\ldots,\mathbf h_K(\mathbf X)]^{\rm H} \in\mathbb C^{K\times MN}.
\label{eq:free_space_channel_matrix}
\end{align}

The signal radiated by each activated element is obtained by multiplying its local guided signal by the corresponding radiation coefficient $q_{m,n}$. This element-wise multiplication is represented by the diagonal radiation coefficient matrix, which is defined as follows:
\begin{equation}
 \mathbf Q(\mathbf Z)=\operatorname{diag}(q_{1,1},\ldots,q_{1,N},q_{2,1},\ldots,q_{M,N})
 \in\mathbb C^{MN\times MN}.
\label{eq:radiation_coefficient_matrix}
\end{equation}
Combining in-waveguide propagation, the radiation coefficients of the activated elements, and free-space propagation yields the effective channel matrix as follows:
\begin{equation}
 \widetilde{\mathbf H}(\mathbf X,\mathbf Z)
 =\mathbf H(\mathbf X)\mathbf Q(\mathbf Z)\mathbf P(\mathbf X)
 \in\mathbb C^{K\times M}.
\label{eq:effective_channel_matrix}
\end{equation}
Its $(k,m)$th entry denotes the effective channel from the $m$th waveguide input to the $k$th user, which is given by
\begin{equation}
 \widetilde h_{k,m}(\mathbf X,\mathbf Z)
 =\sum_{n=1}^{N}c_{k,m,n}q_{m,n}.
\label{eq:effective_waveguide_channel}
\end{equation}
Here, $c_{k,m,n}\triangleq h_{k,m,n}p_{m,n}$ is the cascaded propagation coefficient through the $n$th activated element. The $k$th row of $\widetilde{\mathbf H}$ is denoted by $\widetilde{\mathbf h}_k^{\rm H}=[\widetilde h_{k,1},\ldots,\widetilde h_{k,M}]$. 
\vspace{-5pt}
\subsection{Transmission Model and Problem Formulation}
\label{subsec:transmission}

Let $s_k$ denote the information symbol intended for the $k$th user, and define $\mathbf s=[s_1,\ldots,s_K]^{\rm T}$. The symbols are independently generated with $\mathbb{E}[\mathbf s\mathbf s^{\rm H}]=\mathbf I_K$. These symbols are first processed by the digital precoder, which can be expressed as follows:
\begin{equation}
 \mathbf F=[\mathbf f_1,\ldots,\mathbf f_K]\in\mathbb{C}^{M\times K}.
\label{eq:digital_precoder}
\end{equation}
The digitally precoded signals are then upconverted by the RF chains and fed into the corresponding waveguides at $\mathbf r_{m,0}$.
The received signal at the $k$th user can then be given by
\begin{equation}
\begin{aligned}
 y_k
 &=\mathbf h_k^{\rm H}(\mathbf X)\mathbf Q(\mathbf Z)
 \mathbf P(\mathbf X)\mathbf F\mathbf s+n_k\\
 &=\widetilde{\mathbf h}_k^{\rm H}\mathbf f_k s_k
 +\sum_{i\neq k}\widetilde{\mathbf h}_k^{\rm H}\mathbf f_i s_i+n_k,
\end{aligned}
\label{eq:received_signal}
\end{equation}
where $n_k\sim\mathcal{CN}(0,\sigma_k^2)$ is additive white Gaussian noise. Accordingly, the signal-to-interference-plus-noise ratio (SINR) is given by
\begin{equation}
 \gamma_k=
 \frac{|\widetilde{\mathbf h}_k^{\rm H}\mathbf f_k|^2}
 {\sum_{i\neq k}|\widetilde{\mathbf h}_k^{\rm H}\mathbf f_i|^2+\sigma_k^2}.
\label{eq:sinr}
\end{equation}

To reflect user priorities, we assign weight $\omega_k>0$ to the $k$th user.
The WSR is then given as follows:
\begin{equation}
 R_{\rm WSR}=\sum_{k=1}^{K}\omega_k\log_2(1+\gamma_k).
\label{eq:wsr}
\end{equation}
We jointly optimize the digital precoder $\mathbf F$ and pinching beamforming parameters $(\mathbf X,\mathbf Z)$ by solving
\begin{subequations}
\begin{align}
 \mathcal{P}_0:\quad
 \max_{\mathbf F,\mathbf X,\mathbf Z}\quad&
 \sum_{k=1}^{K}\omega_k\log_2(1+\gamma_k)
 \label{eq:p0_objective}\\
 \mathrm{s.t.}\quad&\|\mathbf F\|_{\rm F}^2\leq P_{\rm t},
 \label{eq:p0_power}\\
 &\mathbf X\in\mathcal{X},\label{eq:p0_position}\\
 &|z_{m,n}|=1,\quad m\in\mathcal{M},~n\in\mathcal{N}.
 \label{eq:p0_state}
\end{align}
\label{prob:joint_design}
\end{subequations}
Here, $P_{\rm t}$ is the total power budget at the $M$ waveguide inputs. Compared with conventional PASS, problem \eqref{prob:joint_design} additionally optimizes the radiation coefficients of the activated elements. In particular, setting $z_{m,n}=\mathrm{j}$ for every activated element yields $q_{m,n}=\mathrm{j}$, which recovers conventional PASS as a special case. With the same activation locations and digital precoder, the two architectures then achieve the same WSR. Optimizing over the additional radiation coefficient variables therefore cannot reduce the maximum WSR. Consequently, the globally optimal WSRs of DM-GPA and conventional PASS, denoted by $R_{\rm D}^{\star}$ and $R_{\rm P}^{\star}$, respectively, satisfy $R_{\rm D}^{\star}\geq R_{\rm P}^{\star}$.

\section{Two-User Analytical Characterization}
\label{sec:two_user_analysis}

To gain analytical insights into the performance benefits of DM-GPA, we first consider a simplified two-user scenario. A limited set of discrete candidate locations makes it difficult for conventional PASS to achieve orthogonal user channels through antenna placement alone. We therefore investigate whether DM-GPA can overcome this limitation by adjusting the radiation coefficients at the selected locations. We begin with one activated element per waveguide, i.e., $N=1$. The following lemma establishes that the two architectures achieve the same maximum WSR in this case.

\begin{lemma}
\label{prop:single_element_equivalence}
If each waveguide has only one activated element, i.e., $N=1$, the maximum WSRs of DM-GPA and conventional PASS satisfy $\left.R_{\rm D}^{\star}\right|_{N=1}=\left.R_{\rm P}^{\star}\right|_{N=1}$.
\end{lemma}

\begin{proof}
For $N=1$, define $\mathbf C(\mathbf X)=[c_{k,m,1}]\in\mathbb C^{K\times M}$ and $\mathbf D_q=\operatorname{diag}(q_{1,1},\ldots,q_{M,1})$. The effective channels of DM-GPA and conventional PASS at the same activation locations are, respectively, $\widetilde{\mathbf H}_{\rm D}=\mathbf C(\mathbf X)\mathbf D_q$ and $\widetilde{\mathbf H}_{\rm P}=\mathrm{j}\mathbf C(\mathbf X)$. For any feasible DM-GPA precoder $\mathbf F_{\rm D}$, consider the PASS precoder
\begin{equation}
 \mathbf F_{\rm P}=-\mathrm{j}\mathbf D_q\mathbf F_{\rm D}.
\label{eq:single_element_equivalent_precoder}
\end{equation}
{Since $|q_{m,1}|\leq1$, this precoder satisfies $\|\mathbf F_{\rm P}\|_{\rm F}^2\leq\|\mathbf F_{\rm D}\|_{\rm F}^2\leq P_{\rm t}$ and $\widetilde{\mathbf H}_{\rm P}\mathbf F_{\rm P}=\widetilde{\mathbf H}_{\rm D}\mathbf F_{\rm D}$.} Hence, conventional PASS can reproduce the received signals and WSR of every feasible DM-GPA design. Conversely, DM-GPA includes conventional PASS as the feasible choice $q_{m,1}=\mathrm{j}$, $\forall m$, as discussed above, which proves Lemma~\ref{prop:single_element_equivalence}.
\end{proof}

\vspace{-3pt}
Lemma~\ref{prop:single_element_equivalence} shows that, for $N=1$, radiation coefficient adjustment can be absorbed into the digital precoder and provides no WSR improvement. In fact, only when $N>1$ can the joint adjustment of multiple radiation coefficients reshape each waveguide's radiation pattern and provide additional beamforming gains.

We therefore consider $M=K=2$ with $N=2$ activated elements per waveguide, equal noise powers $\sigma_1^2=\sigma_2^2=\sigma^2$, and equal user weights $\omega_1=\omega_2=1$. For convenience, we define $\Gamma=P_{\rm t}/\sigma^2$.
For the two-user system, the effective DM-GPA channel in \eqref{eq:effective_channel_matrix} can be written as follows:
\begin{equation}
 \widetilde{\mathbf H}(\mathbf X,\mathbf Z)
 =\begin{bmatrix}
  \widetilde h_{1,1}&\widetilde h_{1,2}\\
  \widetilde h_{2,1}&\widetilde h_{2,2}
 \end{bmatrix},
\label{eq:two_user_effective_channel}
\end{equation}
where $\widetilde h_{k,m}=c_{k,m,1}q_{m,1}+c_{k,m,2}q_{m,2}$ follows from \eqref{eq:effective_waveguide_channel}. Setting $q_{m,n}=\mathrm{j}$ for all activated elements gives the conventional PASS channel at the same activation locations:
\begin{equation}
 \widetilde{\mathbf H}_{\rm P}
 =\mathrm{j}\mathbf H(\mathbf X)\mathbf P(\mathbf X),
 \qquad \widetilde h_{k,m}^{\rm P}=\mathrm{j}\sum_{n=1}^{2}c_{k,m,n}.
\label{eq:pass_fixed_position_channel}
\end{equation}

\subsection{Channel Orthogonalization Through Radiation Coefficient Adjustment}
\label{subsec:two_user_diagonalization}
\label{subsec:two_user_correlation_gain}

We first consider whether radiation coefficient adjustment can transform a nonorthogonal conventional PASS channel into an orthogonal DM-GPA channel without changing the activation locations. Let $\widetilde{\mathbf h}_{k,\rm P}^{\rm H}$ denote the $k$th row of $\widetilde{\mathbf H}_{\rm P}$. The channel gains and normalized channel correlation of conventional PASS are defined as follows:
\begin{equation}
 A_k=\|\widetilde{\mathbf h}_{k,\rm P}\|_2^2,\qquad
 \rho_{\rm P}=\frac{|\widetilde{\mathbf h}_{1,\rm P}^{\rm H}\widetilde{\mathbf h}_{2,\rm P}|}{\sqrt{A_1A_2}},
\label{eq:two_user_pass_correlation}
\end{equation}
where $A_1,A_2>0$. The PASS user channels are orthogonal when $\rho_{\rm P}=0$ and nonorthogonal when $\rho_{\rm P}>0$. Moreover, $|\det\widetilde{\mathbf H}_{\rm P}|^2=A_1A_2(1-\rho_{\rm P}^2)$, so the channel is full rank for $0\leq\rho_{\rm P}<1$ and rank deficient for $\rho_{\rm P}=1$.

To orthogonalize the user channels, we assign the first and second waveguides to the first and second users, respectively, and seek a diagonal DM-GPA channel satisfying $\widetilde h_{1,2}=\widetilde h_{2,1}=0$ and $\widetilde h_{1,1}\widetilde h_{2,2}\neq0$. For the $m$th waveguide, the two activated elements must cancel their signal contributions at the unintended user $k$, where $k\in\{1,2\}$ and $k\neq m$. The required condition is
\begin{equation}
 c_{k,m,1}q_{m,1}+c_{k,m,2}q_{m,2}=0.
\label{eq:two_user_analog_cancellation}
\end{equation}
For a pair of nonzero radiation coefficients, this condition fixes their ratio as
\begin{equation}
 \frac{q_{m,2}}{q_{m,1}}
 =-\frac{c_{k,m,1}}{c_{k,m,2}}
 =\varrho_m e^{\mathrm{j}\delta_m},\qquad
 \delta_m\in(-\pi,\pi],
\label{eq:waveguide_null_ratio}
\end{equation}
where $\varrho_m>0$. Combining this ratio with the Lorentzian constraint yields the closed-form radiation coefficients below. To characterize the desired signal, define $D_m=c_{m,m,1}c_{k,m,2}-c_{m,m,2}c_{k,m,1}$. The condition $D_m\neq0$ ensures that cancelling the unintended-user signal does not also cancel the desired signal.

\begin{lemma}
\label{prop:exact_channel_diagonalization}
Suppose that $0<|\delta_m|<\pi$ and $D_m\neq0$ for both waveguides. {For each waveguide, the Lorentzian-constrained radiation coefficients are given by}
\begin{subequations}
\begin{align}
 \phi_{m,1}^{\rm D}
 &=\operatorname{arccot}_{(0,\pi)}
 \left(\frac{\varrho_m-\cos\delta_m}{\sin\delta_m}\right),
 \label{eq:waveguide_phi_one}\\
 \phi_{m,2}^{\rm D}&=\phi_{m,1}^{\rm D}+\delta_m,
 \label{eq:waveguide_phi_two}\\
 q_{m,n}^{\rm D}&=\sin\phi_{m,n}^{\rm D} e^{\mathrm{j}\phi_{m,n}^{\rm D}},
 \qquad n\in\{1,2\}.
 \label{eq:waveguide_closed_form_state}
\end{align}
\end{subequations}
These radiation coefficients yield $\widetilde{\mathbf H}_{\rm D}=\operatorname{diag}(\widetilde h_{1,1}^{\rm D},\widetilde h_{2,2}^{\rm D})$ with two orthogonal user channels. The resulting effective channel gains are given by
\begin{equation}
 S_m^{\rm D}\triangleq|\widetilde h_{m,m}^{\rm D}|^2
 =\frac{\kappa_m^2|D_m|^2}{|c_{k,m,2}|^2},
\label{eq:retained_desired_gain}
\end{equation}
where
\begin{equation}
 \kappa_m=\frac{|\sin\delta_m|}
 {\sqrt{1+\varrho_m^2-2\varrho_m\cos\delta_m}}.
\label{eq:waveguide_kappa}
\end{equation}
\end{lemma}

\begin{proof}
See Appendix~\ref{app:analog_cancellation} for more details.
\end{proof}

For the resulting diagonal channel, an optimal digital precoder can be chosen as follows:
\begin{equation}
 \mathbf F=\operatorname{diag}(\sqrt{P_1},\sqrt{P_2}),
\label{eq:two_user_diagonal_precoder}
\end{equation}
where $P_1\geq 0$, $P_2\geq 0$, and $P_1+P_2\leq P_{\rm t}$. Under this design, DM-GPA eliminates inter-user interference through radiation coefficient adjustment at fixed element locations. By decoupling interference cancellation from power allocation, the proposed design simplifies digital precoding for multiuser transmission.

However, the Lorentzian constraint can reduce the radiation amplitudes during phase adjustment and weaken the desired signals. We next analyze when the benefit of orthogonalization outweighs this channel power loss. At the given activation locations, let $R_{\rm D}^{\star}(\mathbf X)$ denote the maximum sum rate obtained by jointly optimizing the radiation coefficients and digital precoder, and let $R_{\rm P}^{\star}(\mathbf X)$ denote the maximum conventional PASS sum rate obtained by optimizing the digital precoder with $q_{m,n}=\mathrm{j}$. We adopt the closed-form radiation coefficients and allocate equal power to the two users, i.e., $\mathbf F=\sqrt{P_{\rm t}/2}\mathbf I_2$. The resulting achievable DM-GPA sum rate is
\begin{equation}
 R_{\rm D}^{\perp}=\sum_{m=1}^{2}\log_2\left(1+\frac{\Gamma S_m^{\rm D}}{2}\right)
 \leq R_{\rm D}^{\star}(\mathbf X).
\label{eq:two_user_analog_sum_rate}
\end{equation}
For a feasible digital precoder $\mathbf F$, let $R_{\rm P}(\mathbf F,\mathbf X)$ denote the conventional PASS sum rate obtained from \eqref{eq:wsr}. Allowing the users to jointly process their received signals gives the upper bound~\cite{tse2005fundamentals}
\begin{equation}
\begin{aligned}
 R_{\rm P}(\mathbf F,\mathbf X)
 &\leq\log_2\det\Bigl(\mathbf I_2+\sigma^{-2}\widetilde{\mathbf H}_{\rm P}\mathbf F\mathbf F^{\rm H}\widetilde{\mathbf H}_{\rm P}^{\rm H}\Bigr).
\end{aligned}
\label{eq:two_user_cooperative_bound}
\end{equation}
For $0\leq\rho_{\rm P}<1$, this gives $R_{\rm P}^{\star}(\mathbf X)\leq U_{\rm P}(\mathbf X)$, where
\begin{equation}
\begin{aligned}
 U_{\rm P}(\mathbf X)=\log_2\Bigl[1+\Gamma(A_1+A_2)+\frac{\Gamma^2}{4}A_1A_2(1-\rho_{\rm P}^2)\Bigr].
\end{aligned}
\label{eq:two_user_pass_rate_upper_bound}
\end{equation}
Subtracting this upper bound from $R_{\rm D}^{\perp}$ yields $\mathcal L(\Gamma,\mathbf X)=R_{\rm D}^{\perp}-U_{\rm P}(\mathbf X)$. {The following theorem expresses the rate gain lower bound in terms of the resulting effective channel gains and the original PASS channel correlation.}

\begin{theorem}
\label{prop:two_user_rate_bound}
Based on the orthogonalization design in Lemma~\ref{prop:exact_channel_diagonalization}, a lower bound on the sum-rate gain of DM-GPA over conventional PASS is given by
\begin{equation}
\begin{aligned}
 &\mathcal L(\Gamma,\mathbf X)\!=\!\log_2\!\left[
 \frac{(1+\Gamma S_1^{\rm D}/2)(1+\Gamma S_2^{\rm D}/2)}
 {1+\Gamma(A_1\!+\!A_2)+\frac{\Gamma^2}{4}A_1A_2(1\!-\!\rho_{\rm P}^2)}
 \right].
\end{aligned}
\label{eq:two_user_rate_gain_lower_bound}
\end{equation}
The optimized rates satisfy
\begin{equation}
 R_{\rm D}^{\star}(\mathbf X)-R_{\rm P}^{\star}(\mathbf X)
 \geq\max\{0,\mathcal L(\Gamma,\mathbf X)\}.
\label{eq:two_user_fixed_location_rate_gap}
\end{equation}
\end{theorem}

\begin{proof}
See Appendix~\ref{app:rate_bound_proof} for more details.
\end{proof}

When $\mathcal L\leq0$, the bound in \eqref{eq:two_user_fixed_location_rate_gap} does not guarantee that the benefit of orthogonalization outweighs the channel power loss caused by the Lorentzian constraint. DM-GPA can still match conventional PASS by adopting the full-radiation configuration $q_{m,n}=\mathrm{j}$, without enforcing channel orthogonality. We next identify a sufficient condition for $\mathcal L>0$ at high transmit powers and characterize the asymptotic lower bound.

\begin{lemma}
\label{cor:two_user_correlation_gain}
Consider the orthogonalization design in Lemma~\ref{prop:exact_channel_diagonalization} and suppose that $0\leq\rho_{\rm P}<1$. If $S_1^{\rm D}S_2^{\rm D}>A_1A_2(1-\rho_{\rm P}^2)$, the closed-form DM-GPA radiation coefficients guarantee a sum rate exceeding the maximum attainable by conventional PASS for all sufficiently large transmit powers. As $\Gamma\to\infty$, the rate gain lower bound converges to
\begin{equation}
 \mathcal L_{\infty}=\log_2\frac{S_1^{\rm D}S_2^{\rm D}}{A_1A_2}
 -\log_2(1-\rho_{\rm P}^2).
\label{eq:two_user_asymptotic_rate_gain}
\end{equation}
Moreover, for fixed $A_1,A_2,S_1^{\rm D},S_2^{\rm D}>0$ and $\Gamma>0$, the rate gain lower bound strictly increases with $\rho_{\rm P}$ on $(0,1)$:
\begin{equation}
\begin{aligned}
 &\frac{\partial\mathcal L}{\partial\rho_{\rm P}}=\frac{\Gamma^2A_1A_2\rho_{\rm P}}
 {2\ln2\!\left[1+\Gamma(A_1+A_2)+\frac{\Gamma^2}{4}A_1A_2(1-\rho_{\rm P}^2)\right]}\!>0,
\end{aligned}
\label{eq:two_user_correlation_gain_derivative}
\end{equation}
which indicates that DM-GPA is particularly advantageous when the conventional PASS user channels are strongly correlated.
\end{lemma}

\begin{proof}
See Appendix~\ref{app:correlation_gain_proof} for more details.
\end{proof}

The above analysis demonstrates that DM-GPA can achieve orthogonal user channels by adjusting the radiation coefficients at fixed element locations. The derived lower bound further establishes sufficient conditions for the resulting interference suppression to compensate for the channel power loss caused by the Lorentzian constraint. When $\rho_{\rm P}=1$, the same orthogonalization design can also restore the spatial degree of freedom lost by conventional PASS. We next examine the resulting spatial multiplexing and transmit diversity.

\subsection{Spatial Multiplexing and Transmit Diversity}
\label{subsec:two_user_rate_gain}
\label{subsec:two_user_rank_one_gain}

The use of multiple waveguides in conventional PASS does not necessarily guarantee linearly independent user channels. For certain user and antenna configurations, the effective channel becomes rank deficient, i.e., $\rho_{\rm P}=1$, and supports only one spatial stream. The orthogonalization design in Lemma~\ref{prop:exact_channel_diagonalization} enables DM-GPA to transmit two independent spatial streams at the same element locations.

For conventional PASS, the cooperative upper bound in \eqref{eq:two_user_cooperative_bound} reduces to
\begin{equation}
 R_{\rm P}^{\star}(\mathbf X)\leq\log_2\bigl(1+\Gamma(A_1+A_2)\bigr).
\label{eq:two_user_rank_one_pass_bound}
\end{equation}
Comparing this bound with the achievable DM-GPA rate in \eqref{eq:two_user_analog_sum_rate} yields the following sufficient transmit-power threshold for a strict sum-rate improvement over conventional PASS.

\begin{theorem}
\label{prop:two_user_rank_one_gain}
\label{cor:rank_one_rate_advantage}
Consider the orthogonalization design in Lemma~\ref{prop:exact_channel_diagonalization} and suppose that $\rho_{\rm P}=1$. Then DM-GPA recovers spatial multiplexing and necessarily achieves a strict rate advantage over conventional PASS at sufficiently large transmit power. Specifically, the constructed design satisfies $R_{\rm D}^{\perp}>R_{\rm P}^{\star}(\mathbf X)$ for every $P_{\rm t}>P_{\rm th}$, where a finite sufficient threshold is
\begin{equation}
 P_{\rm th}\triangleq\sigma^2
 \frac{[4(A_1+A_2)-2(S_1^{\rm D}+S_2^{\rm D})]_+}{S_1^{\rm D}S_2^{\rm D}}.
\label{eq:two_user_rank_one_power_threshold}
\end{equation}
If $4(A_1+A_2)-2(S_1^{\rm D}+S_2^{\rm D})\leq0$, then $P_{\rm th}=0$ and the strict rate advantage holds for every positive transmit power.
\end{theorem}

\begin{proof}
See Appendix~\ref{app:rank_one_gain_proof} for more details.
\end{proof}

{Theorem~\ref{prop:two_user_rank_one_gain} shows that, when conventional PASS yields a rank-deficient effective channel, the proposed orthogonalization design guarantees a strict sum-rate advantage at sufficiently high transmit powers, without the additional condition on the channel gains in Lemma~\ref{cor:two_user_correlation_gain}.}

In addition to spatial multiplexing, DM-GPA can achieve transmit diversity by jointly adjusting the radiation coefficients and digital precoder for coherent combining. Both waveguides transmit the same data stream to a single user, and the resulting gain is quantified by the maximum receive SNR.

For the user $k$, define $\mathbf c_{k,m}=[c_{k,m,1},c_{k,m,2}]^{\rm T}$ and $\mathbf q_m=[q_{m,1},q_{m,2}]^{\rm T}$. The effective single-user channel is
\begin{equation}
 \widetilde{\mathbf h}_k^{\rm H}
 =\begin{bmatrix}
 \mathbf c_{k,1}^{\rm T}\mathbf q_1&
 \mathbf c_{k,2}^{\rm T}\mathbf q_2
 \end{bmatrix}.
\label{eq:single_stream_effective_channel}
\end{equation}
Let $\mathbf f\in\mathbb C^2$ denote the digital precoder for this stream, with $\|\mathbf f\|_2^2\leq P_{\rm t}$. At the fixed activation locations, the maximum receive SNR of DM-GPA is
\begin{equation}
 \gamma_{k,\rm D}^{\star}
 =\max_{\substack{\mathbf q_1,\mathbf q_2,\mathbf f\\
 q_{m,n}\in\mathcal Q_{\rm L},\ \|\mathbf f\|_2^2\leq P_{\rm t}}}
 \frac{|\widetilde{\mathbf h}_k^{\rm H}\mathbf f|^2}{\sigma^2}.
\label{eq:single_stream_snr_problem}
\end{equation}
For subsequent characterization, define
\begin{equation}
 L_m=\sum_{n=1}^{2}|c_{k,m,n}|,
 \qquad
 B_m=\left|\sum_{n=1}^{2}c_{k,m,n}\right|.
\label{eq:single_stream_LB}
\end{equation}

\begin{theorem}
\label{prop:single_stream_beamforming_gain}
At the same element locations, DM-GPA and conventional PASS achieve the following maximum receive SNRs with optimal digital precoding:
\begin{subequations}
\begin{align}
 \gamma_{k,\rm D}^{\star}
 &=\frac{\Gamma}{4}\sum_{m=1}^{2}(L_m+B_m)^2,
 \label{eq:single_stream_dma_snr}\\
 \gamma_{k,\rm P}^{\star}
 &=\Gamma\sum_{m=1}^{2}B_m^2.
 \label{eq:single_stream_pass_snr}
\end{align}
\end{subequations}
The resulting improvement over conventional PASS is
\begin{equation}
 \gamma_{k,\rm D}^{\star}-\gamma_{k,\rm P}^{\star}
 =\frac{\Gamma}{4}\sum_{m=1}^{2}
 (L_m-B_m)(L_m+3B_m)\geq0.
\label{eq:single_stream_snr_gain}
\end{equation}
\end{theorem}

\begin{proof}
For fixed radiation coefficients, maximum-ratio transmission is optimal, and \eqref{eq:single_stream_snr_problem} reduces to
\begin{equation}
 \gamma_{k,\rm D}^{\star}
 =\Gamma\max_{\mathbf q_1,\mathbf q_2}
 \sum_{m=1}^{2}|\mathbf c_{k,m}^{\rm T}\mathbf q_m|^2.
\label{eq:single_stream_mrt_reduction}
\end{equation}
Using $q_{m,n}=(\mathrm{j}+z_{m,n})/2$ with $|z_{m,n}|=1$, the triangle inequality gives
\begin{equation}
 |\mathbf c_{k,m}^{\rm T}\mathbf q_m|
 \leq\frac{1}{2}(B_m+L_m).
\label{eq:single_stream_waveguide_bound}
\end{equation}
When $B_m>0$, this bound is attained by choosing the phases of $z_{m,1}$ and $z_{m,2}$ such that the two adjustable terms align with the fixed term $\mathrm{j}\sum_{n=1}^{2}c_{k,m,n}$. When $B_m=0$, the adjustable terms can instead be aligned with each other, which also attains the bound. Since the radiation coefficients are independently adjustable across waveguides, substituting the attained bound into \eqref{eq:single_stream_mrt_reduction} proves \eqref{eq:single_stream_dma_snr}. For conventional PASS, $q_{m,n}=\mathrm{j}$ gives $|\mathbf c_{k,m}^{\rm T}\mathbf q_m|=B_m$, which proves \eqref{eq:single_stream_pass_snr}. Expanding their difference gives \eqref{eq:single_stream_snr_gain}. Finally, $B_m\leq L_m$ by the triangle inequality, with strict inequality when the two nonzero PASS radiation contributions on the $m$th waveguide are not phase aligned.
\end{proof}

The two-user analysis establishes that radiation coefficient adjustment provides DM-GPA with additional control over the effective user channels at fixed element locations. The derived sum-rate and transmit diversity results demonstrate that this control can yield performance improvements over conventional PASS despite the Lorentzian constraint. We next exploit this design freedom in general multiuser transmission through joint pinching beamforming and digital precoder design.
\section{Weighted Sum Rate Maximization}
\label{sec:joint_optimization}

The WSR maximization problem in \eqref{prob:joint_design} is nonconvex due to the coupling between pinching beamforming and the digital precoder, the discrete element locations, and the Lorentzian constraint. We develop an AO algorithm based on an equivalent WMMSE formulation to alternately update pinching beamforming and the digital precoder. For pinching beamforming, a sequential element-wise search optimizes the locations of the activated elements, with a closed-form radiation coefficient solution at each candidate location.
\subsection{WMMSE Reformulation and Digital Precoder Design}
\label{subsec:wmmse}

\subsubsection{{Power-Normalized WMMSE Reformulation}}
\label{subsec:wmmse_equivalence}

{
To simplify the digital precoder optimization, we first remove the explicit transmit power constraint through power normalization. The following lemma shows that the optimal digital precoder uses the full transmit power budget.

\begin{lemma}
\label{lemma:full_power}
For any fixed feasible $(\mathbf X,\mathbf Z)$ with at least one nonzero user channel, the optimal digital precoder satisfies
\begin{equation}
 \|\mathbf F^\star\|_{\rm F}^2=P_{\rm t}.
\label{eq:full_power}
\end{equation}
\end{lemma}

\begin{proof}
Suppose that an optimal precoder uses power $P<P_{\rm t}$. Scaling every precoding vector by $\sqrt{P_{\rm t}/P}$ increases each desired-signal power and its corresponding interference powers by the same factor, while the receiver noise remains unchanged. Consequently, this scaling strictly increases the SINR of every user with a nonzero desired signal and leaves the other users' SINRs unchanged. The resulting increase in WSR contradicts the assumed optimality of the precoder.
\end{proof}

Following the variational power normalization in~\cite{bereyhi2026mimopass}, consider an unconstrained precoder $\overline{\mathbf F}=[\overline{\mathbf f}_1,\ldots,\overline{\mathbf f}_K]$ and define
\begin{equation}
 \overline\gamma_k=
 \frac{|\widetilde{\mathbf h}_k^{\rm H}\overline{\mathbf f}_k|^2}
 {\sum_{i\neq k}|\widetilde{\mathbf h}_k^{\rm H}\overline{\mathbf f}_i|^2
 +\dfrac{\sigma_k^2}{P_{\rm t}}\|\overline{\mathbf F}\|_{\rm F}^2}.
\label{eq:variational_sinr}
\end{equation}
The power-feasible precoder recovered from $\overline{\mathbf F}$ is
\begin{equation}
 \widehat{\mathbf F}
 =\frac{\sqrt{P_{\rm t}}}{\|\overline{\mathbf F}\|_{\rm F}}
 \overline{\mathbf F}.
\label{eq:precoder_rescaling}
\end{equation}
Direct substitution shows that $\gamma_k(\widehat{\mathbf F},\mathbf X,\mathbf Z)=\overline\gamma_k(\overline{\mathbf F},\mathbf X,\mathbf Z)$ for every $k$. Together with Lemma~\ref{lemma:full_power}, this establishes the equivalence between \eqref{prob:joint_design} and the following power-unconstrained problem:
\begin{subequations}
\begin{align}
 \overline{\mathcal P}_0:\quad
 \max_{\overline{\mathbf F}\neq\mathbf0,\mathbf X,\mathbf Z}\quad&
 \sum_{k=1}^{K}\omega_k\log_2(1+\overline\gamma_k)
 \label{eq:variational_objective}\\
 \mathrm{s.t.}\quad&\mathbf X\in\mathcal X,\qquad
 |z_{m,n}|=1,~\forall m,n.
 \label{eq:variational_constraints}
\end{align}
\label{prob:variational_design}
\end{subequations}
Thus, the power budget is incorporated into the effective noise term in \eqref{eq:variational_sinr}. The digital precoder $\widehat{\mathbf F}$ is obtained by the normalization in \eqref{eq:precoder_rescaling}.

We now apply the WSR-WMMSE equivalence~\cite{christensen2008weighted,shi2011wmmse} to \eqref{prob:variational_design}. For notational simplicity, the overbar on the variational precoder is omitted in the remainder of the algorithm derivation. Define $\xi_{k,i}\triangleq\widetilde{\mathbf h}_k^{\rm H}\mathbf f_i$ and
\begin{equation}
 T_k\triangleq\sum_{i=1}^{K}|\xi_{k,i}|^2
 +\frac{\sigma_k^2}{P_{\rm t}}\|\mathbf F\|_{\rm F}^2.
\label{eq:total_received_power}
\end{equation}
The corresponding variational MSE associated with a scalar receive filter $u_k\in\mathbb C$ is defined as follows:
\begin{equation}
\begin{aligned}
 e_k\triangleq|u_k|^2T_k
 -2\operatorname{Re}\{u_k^*\xi_{k,k}\}+1.
\end{aligned}
\label{eq:mse}
\end{equation}

To relate the MSEs to the WSR, introduce a positive auxiliary weight $v_k$ for each MSE. Retaining weights $\omega_k$ in \eqref{eq:wsr}, we define the WMMSE objective as follows:
\begin{equation}
 \mathcal{J}=\sum_{k=1}^{K}\omega_k(v_ke_k-\ln v_k).
\label{eq:wmmse_objective}
\end{equation}
The receive filters and auxiliary MSE weights are collected in $\mathbf u=[u_1,\ldots,u_K]^{\rm T}$ and $\mathbf v=[v_1,\ldots,v_K]^{\rm T}$, respectively. The following lemma establishes the relationship between the minimum of \eqref{eq:wmmse_objective} over these auxiliary variables and the original WSR.

\begin{lemma}
\label{lemma:wmmse_equivalence}
For any fixed $(\mathbf F,\mathbf X,\mathbf Z)$ with $\mathbf F\neq\mathbf0$, the optimal receive filters and MSE weights, together with the resulting MMSE, are given as follows:
\begin{equation}
\begin{aligned}
 u_k^\star&=\frac{\xi_{k,k}}{T_k},\qquad e_k^{\rm MMSE}=\frac{1}{1+\overline\gamma_k},\\
 v_k^\star&=\frac{1}{e_k^{\rm MMSE}}=1+\overline\gamma_k,
\end{aligned}
\label{eq:mmse_total}
\end{equation}
where $\overline\gamma_k$ is the variational SINR in \eqref{eq:variational_sinr}. Moreover,
\begin{equation}
 \min_{\mathbf u,\mathbf v\succ\mathbf 0}\mathcal{J}
 =\sum_{k=1}^{K}\omega_k
 -\ln(2)\sum_{k=1}^{K}\omega_k\log_2(1+\overline\gamma_k).
\label{eq:wmmse_wsr_affine}
\end{equation}
\end{lemma}

\begin{proof}
See Appendix~\ref{app:wmmse_proof} for more details.
\end{proof}

Since the first term on the right-hand side of \eqref{eq:wmmse_wsr_affine} is constant, \eqref{prob:variational_design} is equivalent to
\begin{subequations}
\begin{align}
 \mathcal{P}_1:\quad
 \min_{\mathbf F, \mathbf X,\mathbf Z,\mathbf u,\mathbf v}\quad&\mathcal{J}
 \label{eq:p1_objective}\\
 \mathrm{s.t.}\quad&\mathbf X\in\mathcal{X},\label{eq:p1_resource}\\
 &|z_{m,n}|=1,\quad v_k>0,~\forall m,n,k.
 \label{eq:p1_state}
\end{align}
\label{prob:wmmse_design}
\end{subequations}
We address \eqref{prob:wmmse_design} by alternately updating the auxiliary variables, the digital precoder, and pinching beamforming. With the other variables fixed, the digital precoder update reduces to an unconstrained quadratic problem with a closed-form solution.
}

\subsubsection{{ Digital Precoder Update}}
\label{subsec:digital_precoder}

{
For fixed $(\mathbf X,\mathbf Z,\mathbf u,\mathbf v)$, define
\begin{equation}
 \tau\triangleq\frac{1}{P_{\rm t}}
 \sum_{k=1}^{K}\omega_kv_k|u_k|^2\sigma_k^2.
\label{eq:explicit_regularizer}
\end{equation}
The terms of \eqref{eq:wmmse_objective} that depend on $\mathbf F$ yield the unconstrained quadratic problem
\begin{equation}
\begin{aligned}
 \min_{\mathbf F}\quad&
 \operatorname{Tr}\!\left[\mathbf F^{\rm H}
 (\mathbf A+\tau\mathbf I_M)\mathbf F\right]
 -2\operatorname{Re}\{\operatorname{Tr}(\mathbf D^{\rm H}\mathbf F)\},
\end{aligned}
\label{prob:digital_precoder}
\end{equation}
where
\begin{subequations}
\begin{align}
 \mathbf A&=\sum_{k=1}^{K}\omega_kv_k|u_k|^2
 \widetilde{\mathbf h}_k\widetilde{\mathbf h}_k^{\rm H},
 \label{eq:A_matrix}\\
 \mathbf D&=[\omega_1v_1u_1\widetilde{\mathbf h}_1,\ldots,
 \omega_Kv_Ku_K\widetilde{\mathbf h}_K].
 \label{eq:D_matrix}
\end{align}
\end{subequations}
Since $\tau>0$ for nondegenerate transmission, $\mathbf A+\tau\mathbf I_M$ is positive definite. The unique minimizer is therefore
\begin{equation}
 \mathbf F^\star=(\mathbf A+\tau\mathbf I_M)^{-1}\mathbf D.
\label{eq:precoder_update}
\end{equation}
}

\subsection{Pinching Beamforming Optimization}
\label{subsec:position_state}

We next optimize pinching beamforming using a sequential element-wise search that jointly updates the activation location and Lorentzian-constrained radiation coefficient of each activated element. During each update, the digital precoder $\mathbf F$, the auxiliary variables $(\mathbf u,\mathbf v)$, and the activation locations and radiation coefficients of all remaining activated elements are fixed. Consider the update of the $n$th element on the $m$th waveguide, whose current activation location and radiation coefficient are denoted by $x_{m,n}^{\rm old}$ and $q_{m,n}^{\rm old}$, respectively.

\subsubsection{Removing the current element}
\label{subsec:remove_element}

For a trial activation location $x\in\mathcal{S}_x$, let $c_{k,m}(x)$ denote the cascaded coefficient $c_{k,m,n}$ evaluated at $x_{m,n}=x$. Removing the current element contribution from the effective response $\xi_{k,i}$ gives the residual response as follows:
\begin{equation}
 \overline{\xi}_{k,i}^{(m,n)}
 =\xi_{k,i}-f_{m,i}c_{k,m}(x_{m,n}^{\rm old})q_{m,n}^{\rm old}.
\label{eq:residual_response}
\end{equation}
Using the radiation coefficient parameterization in \eqref{eq:lorentzian_theta}, the updated effective response at the trial location $x$ can be written as follows:
\begin{equation}
\begin{aligned}
 \xi_{k,i}(x,z)
 &=\overline{\xi}_{k,i}^{(m,n)}+f_{m,i}c_{k,m}(x)\frac{\mathrm{j}+z}{2}\\
 &=b_{k,i}(x)+a_{k,i}(x)z,
\end{aligned}
\label{eq:affine_stream_response}
\end{equation}
where $a_{k,i}(x)=f_{m,i}c_{k,m}(x)/2$ and $b_{k,i}(x)=\overline{\xi}_{k,i}^{(m,n)}+\mathrm{j} a_{k,i}(x)$.

\subsubsection{Closed-form radiation coefficient for each trial location}
\label{subsec:closed_form_state}

Substituting \eqref{eq:affine_stream_response} into \eqref{eq:wmmse_objective} and using $|z|=1$ yields
\begin{equation}
 \mathcal{J}_{m,n}(x,z)
 =C_{m,n}(x)+2\operatorname{Re}\{\zeta_{m,n}(x)z\},
\label{eq:coordinate_objective}
\end{equation}
Here, $\bar T_k(x)$ denotes the part of the variational received power in \eqref{eq:total_received_power} that is independent of $z$, given as follows:
\begin{equation}
 \bar T_k(x)\triangleq\sum_{i=1}^{K}
 \big(|b_{k,i}(x)|^2+|a_{k,i}(x)|^2\big)
 +\frac{\sigma_k^2}{P_{\rm t}}\|\mathbf F\|_{\rm F}^2.
\label{eq:state_independent_received_power}
\end{equation}
The term in \eqref{eq:coordinate_objective} that is independent of $z$ takes the form
\begin{equation}
\begin{split}
 C_{m,n}(x)&=\sum_{k=1}^{K}\omega_kv_k\Big[|u_k|^2\bar T_k(x)
 -2\operatorname{Re}\{u_k^*b_{k,k}(x)\}+1\Big]\\
 &-\sum_{k=1}^{K}\omega_k\ln v_k,
\end{split}
\label{eq:C_definition}
\end{equation}
and the coefficient of the term dependent on $z$ is given by
\begin{equation}
 \zeta_{m,n}(x)=\sum_{k=1}^{K}\omega_kv_k
 \Big[|u_k|^2\sum_{i=1}^{K}b_{k,i}^*(x)a_{k,i}(x)
 -u_k^*a_{k,k}(x)\Big].
\label{eq:c_definition}
\end{equation}

\begin{lemma}
\label{prop:closed_form_state}
With $\mathbf F$, $\mathbf u$, $\mathbf v$, and all other elements fixed, the optimal unit-modulus auxiliary variable for a feasible trial location $x$ in the single-element WMMSE subproblem is obtained as follows:
\begin{equation}
 z_{m,n}^{\star}(x)
 =-\frac{\zeta_{m,n}^*(x)}{|\zeta_{m,n}(x)|},
 \qquad \zeta_{m,n}(x)\neq0.
\label{eq:z_closed_form}
\end{equation}
The corresponding Lorentzian-constrained radiation coefficient and minimum WMMSE value after optimization can then be evaluated as follows:
\begin{subequations}
\begin{align}
 q_{m,n}^{\star}(x)&=\frac{\mathrm{j}+z_{m,n}^{\star}(x)}{2},
 \label{eq:q_closed_form}\\
 \mathcal{J}_{m,n}^{\star}(x)&=C_{m,n}(x)-2|\zeta_{m,n}(x)|.
 \label{eq:profiled_score}
\end{align}
\end{subequations}
If $\zeta_{m,n}(x)=0$, every unit-modulus value of $z$ is optimal for the current coordinate subproblem, and its previous value can be retained.
\end{lemma}

\begin{proof}
The coefficient and auxiliary variable can be expressed in polar form as $\zeta_{m,n}(x)=|\zeta_{m,n}(x)|e^{\mathrm{j}\phi}$ and $z=e^{\mathrm{j}\vartheta}$, respectively. The only term in \eqref{eq:coordinate_objective} that depends on $z$ is $2|\zeta_{m,n}(x)|\cos(\phi+\vartheta)$, whose minimum is attained when $\vartheta=\pi-\phi$ mod $2\pi$. This gives \eqref{eq:z_closed_form} and \eqref{eq:profiled_score}. 
\end{proof}
For each feasible activation location, Lemma~\ref{prop:closed_form_state} provides the optimal radiation coefficient and the corresponding minimum WMMSE value. The activation location can then be selected by comparing these values across the feasible candidates.

\subsubsection{Search over feasible activation locations}
\label{subsec:finite_search}

When updating the $n$th element on the $m$th waveguide, the feasible candidate set excludes locations occupied by the other activated elements on that waveguide and is given by
\begin{equation}
 \mathcal{S}_{m,n}^{\rm feas}
 =\mathcal{S}_x\setminus
 \left\{x_{m,r}:r\in\mathcal{N},~r\neq n\right\}.
\label{eq:feasible_candidate_set}
\end{equation}
Substituting the optimal auxiliary variable from Lemma~\ref{prop:closed_form_state} into the objective gives the exact solution of the joint activation location and radiation coefficient subproblem for the activated element as follows:
\begin{subequations}\label{eq:rad_coefficient_update}
\begin{align}
 x_{m,n}^{\star}
 &\in\arg\min_{x\in\mathcal{S}_{m,n}^{\rm feas}}
 \big[C_{m,n}(x)-2|\zeta_{m,n}(x)|\big],
 \label{eq:position_update}\\
 z_{m,n}^{\star}
 &=z_{m,n}^{\star}(x_{m,n}^{\star}).
 \label{eq:state_after_position}
\end{align}
\end{subequations}

For each feasible activation location, we compute the optimal radiation coefficient and the corresponding WMMSE value. The location yielding the lowest WMMSE is then selected together with its associated radiation coefficient, which completes the joint activation location and radiation coefficient update for the current element.

\subsection{Complexity and Convergence Analysis}
\label{subsec:algorithm}

The proposed AO algorithm for solving~\eqref{prob:joint_design} is summarized in Algorithm~\ref{alg:joint_wmmse}.

\begin{algorithm}[t]
\caption{WMMSE-Based Joint Digital Precoder and Pinching Beamforming Optimization}
\label{alg:joint_wmmse}
\begin{algorithmic}[1]
\REQUIRE Candidate channels $\{c_{k,m}(x):x\in\mathcal{S}_x\}$, $P_{\rm t}$,
and $\{\omega_k\}$.
\ENSURE $\mathbf F^\star$, $\mathbf X^\star$, and $\mathbf Z^\star$.
\STATE Set $R_{\rm best}\leftarrow-\infty$.
    \STATE {Initialize feasible $\mathbf X$, $\mathbf F$, and set $z_{m,n}=\mathrm{j}$.}
    \REPEAT
        \STATE {For fixed $(\mathbf X,\mathbf Z)$, iterate the updates in
        \eqref{eq:mmse_total} and \eqref{eq:precoder_update} until convergence.}
        \STATE For fixed $\mathbf F$, perform sequential activation location and radiation coefficient updates using \eqref{eq:rad_coefficient_update}.
        Update $\xi_{k,i}$ after each element and refresh $(\mathbf u,\mathbf v)$
        by \eqref{eq:mmse_total}.
    \UNTIL{the outer WSR converges}
    \STATE {Recover $\widehat{\mathbf F}$ using \eqref{eq:precoder_rescaling}.}
    \STATE {Update $(R_{\rm best},\mathbf F^\star,\mathbf X^\star,\mathbf Z^\star)$ if the
    converged WSR exceeds $R_{\rm best}$, with $\mathbf F^\star\leftarrow\widehat{\mathbf F}$.}
\RETURN $(\mathbf F^\star,\mathbf X^\star,\mathbf Z^\star)$.
\end{algorithmic}
\end{algorithm}

\subsubsection{Complexity analysis}
\label{subsec:complexity}

With precomputed candidate channels, each outer AO iteration involves $I_{\rm F}$ WMMSE iterations for digital precoding and $I_{\rm X}$ sweeps of the sequential element-wise search for pinching beamforming. Each WMMSE iteration requires $\mathcal{O}(M^3+KM^2)$ operations to form the coefficient matrices and solve \eqref{eq:precoder_update}. For pinching beamforming, each candidate evaluation costs $\mathcal{O}(K^2)$, and a sweep over all $MN$ activated elements has complexity $\mathcal{O}(MN\bar Q K^2)$, where $\bar Q$ is the average number of feasible candidate locations per element. For $I_{\rm AO}$ outer iterations, the total complexity per initialization is therefore
\begin{equation}
 \mathcal{O}\!\left[I_{\rm AO}\left(
 I_{\rm F}(M^3+KM^2)+I_{\rm X}MN\bar Q K^2\right)\right].
\label{eq:total_complexity}
\end{equation}
\begin{figure*}[!t]
\centering
\subfloat[Sum rate versus channel correlation.\label{fig:two_user_rate_correlation}]{%
\makebox[0.32\textwidth][c]{\includegraphics[height=0.21\textwidth]{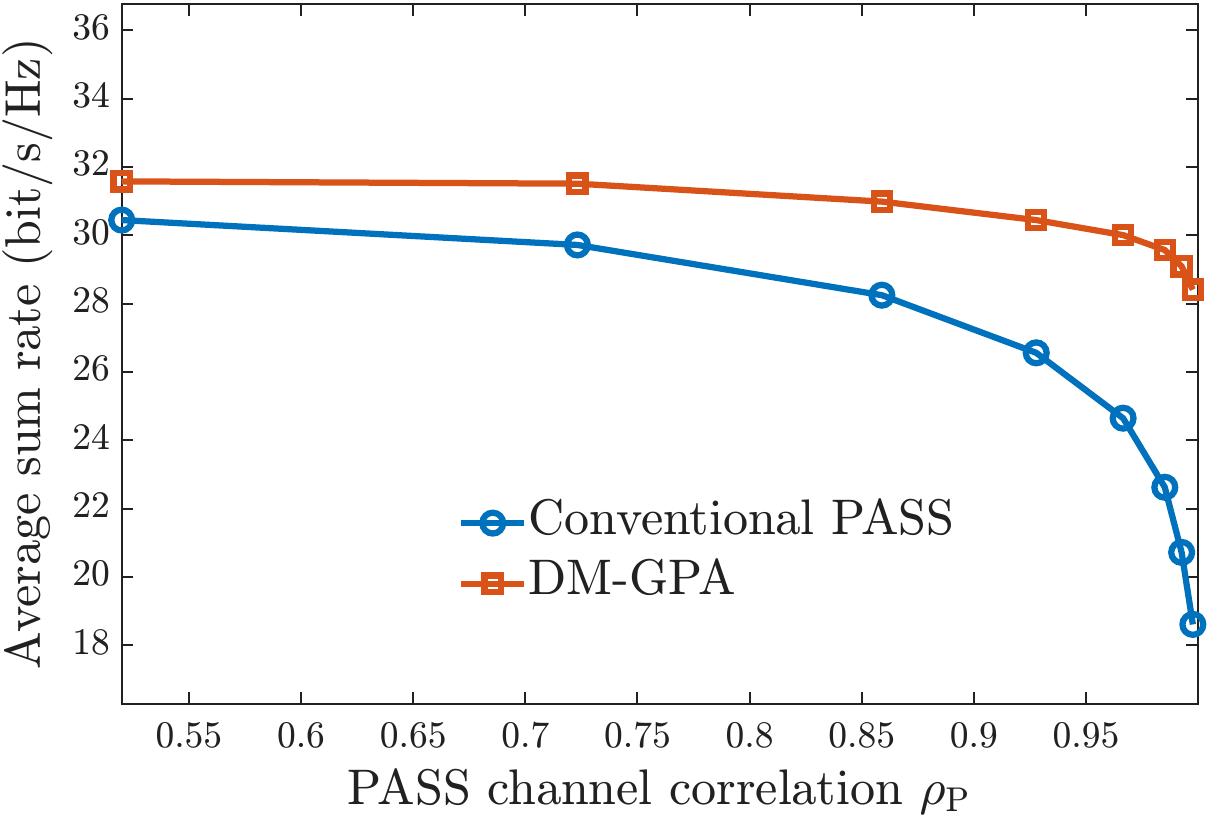}}}%
\hfill
\subfloat[PASS channel correlation versus the number of candidate locations.\label{fig:two_user_correlation_candidates}]{%
\makebox[0.32\textwidth][c]{\includegraphics[height=0.21\textwidth]{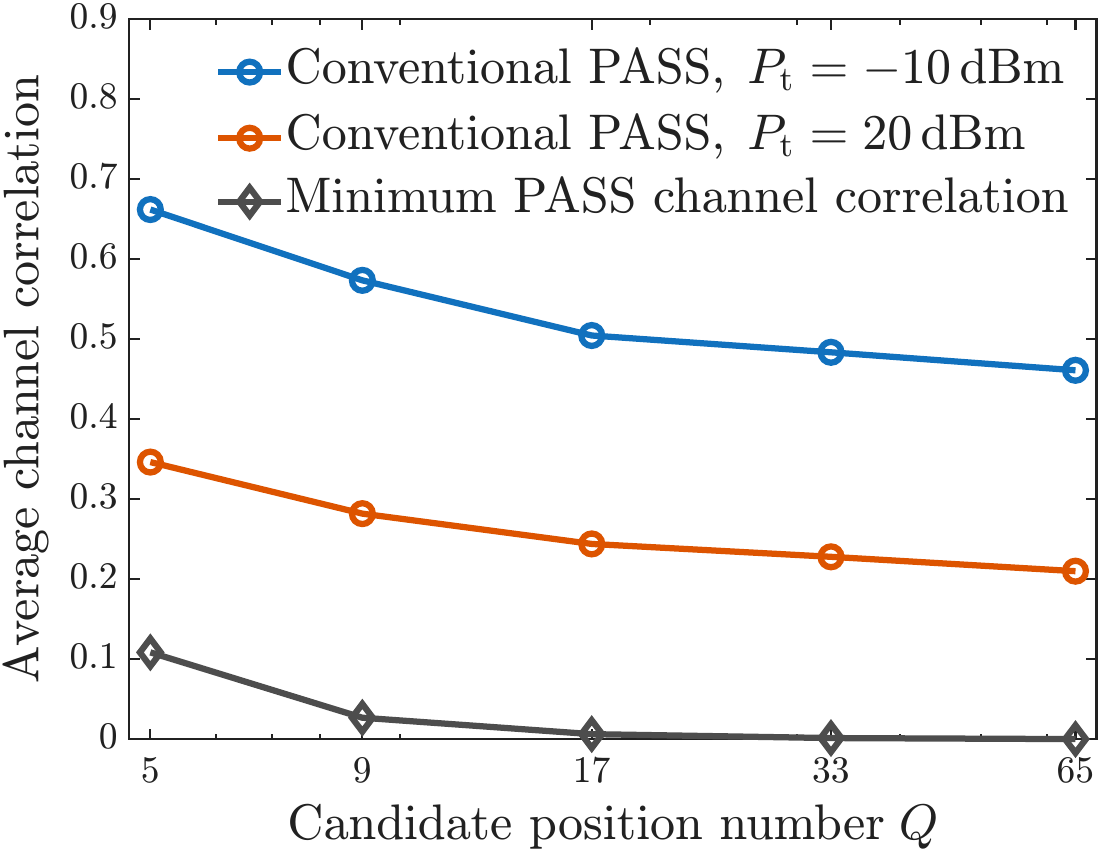}}}%
\hfill
\subfloat[Sum rate versus transmit power.\label{fig:two_user_rate_power}]{%
\makebox[0.32\textwidth][c]{\includegraphics[height=0.21\textwidth]{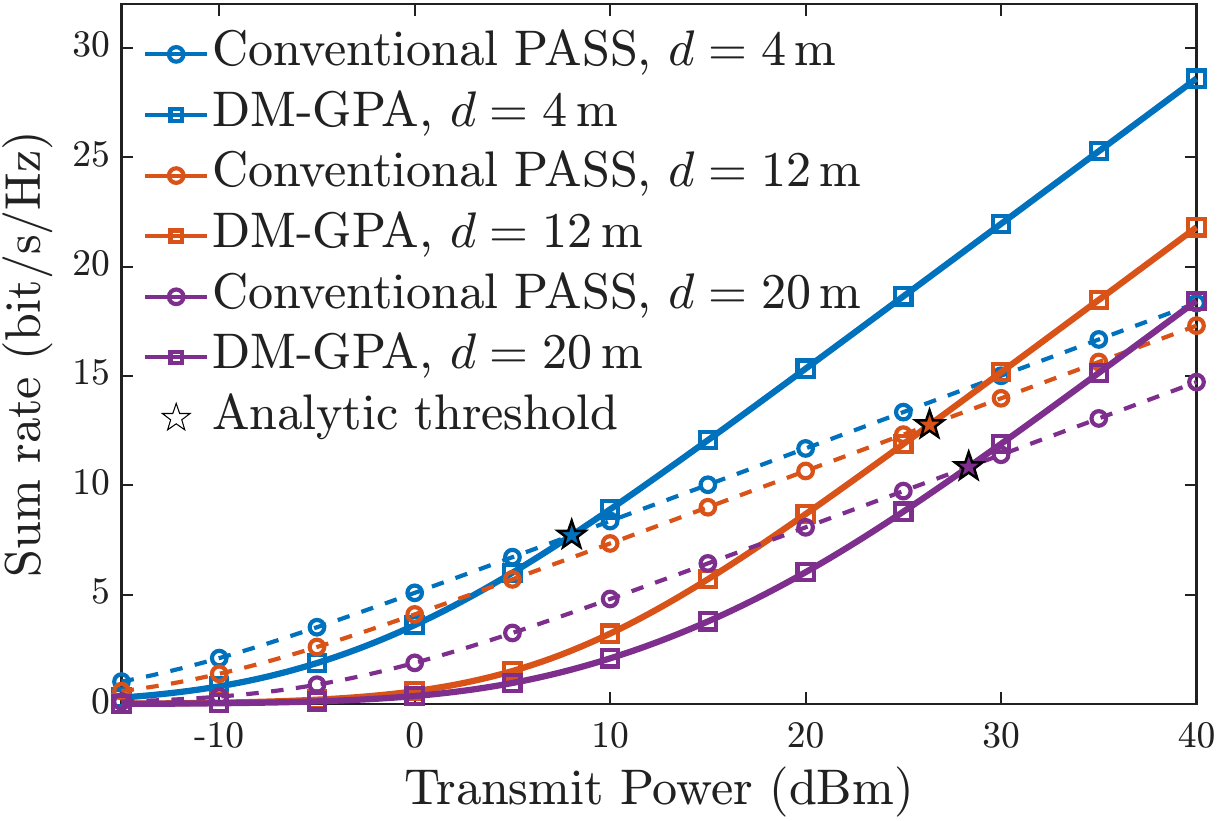}}}
\caption{Two-user performance comparison of DM-GPA and conventional PASS.}
\label{fig:two_user_performance}\vspace{-10pt}
\end{figure*}
\subsubsection{Convergence analysis}
\label{subsec:convergence}

{At each iteration, the auxiliary-variable updates, the closed-form digital precoder update, and the element-wise pinching beamforming search do not increase the variational WMMSE objective in~\eqref{eq:wmmse_objective}. By~\eqref{eq:wmmse_wsr_affine}, the corresponding variational WSR is monotonically nondecreasing. Equation~\eqref{eq:precoder_rescaling} maps every nonzero variational precoder to the feasible digital precoder $\widehat{\mathbf F}$ with the same user SINRs. The variational WSR is therefore bounded above by the finite optimum of \eqref{prob:joint_design}, which proves convergence of the objective-value sequence generated by Algorithm~\ref{alg:joint_wmmse}.}

\section{Numerical Results}
\label{sec:numerical_results}

We now validate the effectiveness of the proposed DM-GPA framework and optimization algorithm through numerical simulations. Specifically, we compare DM-GPA with conventional PASS, LCX-GPA, and fixed-array baselines. The evaluation examines algorithm convergence and the WSR performance of DM-GPA under different system configurations.

\subsection{Simulation Setup}
\label{subsec:simulation_setup}

Unless otherwise specified, we set $f_c=28$ GHz, $n_{\rm eff}=1.44$, $\epsilon=0.66$ dB/m~\cite{epp2005hardware}, $P_{\rm t}=0$ dBm, and the noise power at each user to $-90$ dBm. The waveguides are uniformly spaced across $D_y=5$ m at height $h=3$ m, with uniformly spaced candidate locations. Users are independently and uniformly distributed over the $D_x\times D_y$ m$^2$ ground region, with $\omega_k=1$. Other configuration parameters are given in the figure captions.

For each realization, DM-GPA starts from random feasible activation locations, radiation coefficients $q_{m,n}=\mathrm{j}$, and a power-normalized regularized zero-forcing (RZF) precoder given as follows:
\begin{subequations}
\begin{align}
 \widetilde{\mathbf F}^{(0)}
 &=\widetilde{\mathbf H}^{\rm H}
 (\widetilde{\mathbf H}\widetilde{\mathbf H}^{\rm H}+\alpha_{\rm RZF}\mathbf I_K)^{-1},
 \label{eq:rzf}\\
 \mathbf F^{(0)}
 &=\sqrt{P_{\rm t}}\frac{\widetilde{\mathbf F}^{(0)}}
 {\|\widetilde{\mathbf F}^{(0)}\|_{\rm F}}.
 \label{eq:rzf_normalization}
\end{align}
\end{subequations}
Here, $\alpha_{\rm RZF}$ is the regularization parameter. Reconfigurable baselines use feasible initial locations under their respective propagation models. The outer AO iterations stop when the relative WSR improvement falls below $\varepsilon_{\rm AO}=10^{-4}$. Unless otherwise stated, general multiuser results are averaged over $1000$ independent user deployments.

\subsection{Baseline Architectures}
\label{subsec:benchmark_architectures}

We consider the following four baselines under the same transmit power constraint:
\begin{enumerate}[label=(\arabic*),leftmargin=*,nosep]
 \item \emph{\bf Conventional PASS:} This scheme uses the DM-GPA deployment, candidate set, number of activated elements, and in-waveguide channel model, with fixed coefficients $q_{m,n}=\mathrm{j}$. Only activation locations and the $M\times K$ digital precoder are alternately optimized.

 \item \emph{\bf LCX-GPA:} Leaky coaxial cables replace the waveguides, with relative permittivity $\epsilon_r=1.26$, propagation constant $\beta_{\rm LCX}=k_0\sqrt{\epsilon_r}$~\cite{wang2025lcxgpa}, and the same attenuation coefficient as DM-GPA. The channel includes direction-dependent leakage. Slot activation locations and the $M\times K$ digital precoder are jointly optimized.

 \item \emph{\bf Fixed hybrid array:} An $M\times N$ uniform planar array with half-wavelength spacing is centered at $\mathbf r_{\rm f}=[0,0,h]^{\rm T}$ on the waveguide-feed side. Its element coordinates are given as follows:
 \begin{equation}
  \mathbf r_{m,n}^{\rm fix}=\mathbf r_{\rm f}
  +\frac{\lambda}{2}
  \begin{bmatrix}
   n-(N+1)/2\\ m-(M+1)/2\\ 0
  \end{bmatrix}.
 \label{eq:fixed_array_coordinates}
 \end{equation}
 Each of the $M$ RF chains drives $N$ elements through constant-modulus quantized phase shifters in a sub-connected structure. The $M\times K$ digital precoder is optimized by WMMSE.

 \item \emph{\bf Fixed digital array:} This array uses the coordinates in \eqref{eq:fixed_array_coordinates} and one RF chain per element. Its $MN\times K$ digital precoder is optimized by WMMSE. 
\end{enumerate}

\subsection{Performance Evaluation in the Two-User Scenario}
\label{subsec:numerical_two_user}

We first consider a two-user scenario to validate the analytical results in Section~\ref{sec:two_user_analysis}. Fig.~\ref{fig:two_user_rate_correlation} examines the effect of PASS channel correlation on the average sum rates of DM-GPA and conventional PASS. As can be observed, the sum rate of conventional PASS decreases sharply as $\rho_{\rm P}$ approaches one. In contrast, DM-GPA experiences a smaller rate loss because radiation coefficient adjustment provides additional flexibility to suppress inter-user interference. As a result, its sum-rate advantage becomes more pronounced for strongly correlated PASS channels. This trend is consistent with Lemma~\ref{cor:two_user_correlation_gain}.

To relate this advantage to the available candidate locations, Fig.~\ref{fig:two_user_correlation_candidates} further shows the PASS channel correlation after antenna location optimization for different values of $Q$. As shown in the figure, the correlation decreases with $Q$ because more candidate locations provide greater flexibility to separate the user channels. Another important observation is that the optimized correlation is lower at the higher transmit power. This is because inter-user interference has a greater impact on the sum rate at high transmit power, whereas stronger channel gains are more important at low transmit power. Nevertheless, the optimized correlations remain above the minimum achievable values because correlation minimization alone does not account for channel gains. Taken together, Figs.~\ref{fig:two_user_rate_correlation} and~\ref{fig:two_user_correlation_candidates} indicate that DM-GPA can offer a more pronounced sum-rate advantage when limited candidate locations leave conventional PASS with strongly correlated user channels.

We further consider the rank-deficient case $\rho_{\rm P}=1$ to examine the spatial multiplexing advantage of DM-GPA. Fig.~\ref{fig:two_user_rate_power} evaluates the closed-form orthogonalization design against conventional PASS for different inter-waveguide spacings $d$. As shown in the figure, this design yields a lower sum rate at low transmit power because the channel power loss caused by the Lorentzian constraint outweighs the benefit of interference cancellation. At high transmit power, however, DM-GPA achieves a faster rate increase because orthogonalization enables two independent spatial streams, whereas conventional PASS supports only one. As a result, DM-GPA achieves a strictly higher sum rate above the analytical transmit-power thresholds for all considered spacings, in agreement with Theorem~\ref{prop:two_user_rank_one_gain}. These results confirm that the recovered spatial multiplexing can compensate for the channel power loss under the Lorentzian constraint.

\subsection{Performance Evaluation in General Multiuser Scenarios}
\label{subsec:numerical_multiuser}
We next evaluate the convergence of the proposed algorithm and compare the performance of DM-GPA with the considered benchmark schemes in general multiuser scenarios.

\subsubsection{Convergence of the Joint Optimization Algorithm}
\label{subsec:numerical_ao_convergence}

\begin{figure}[!t]
\centering
\includegraphics[width=0.7\columnwidth]
{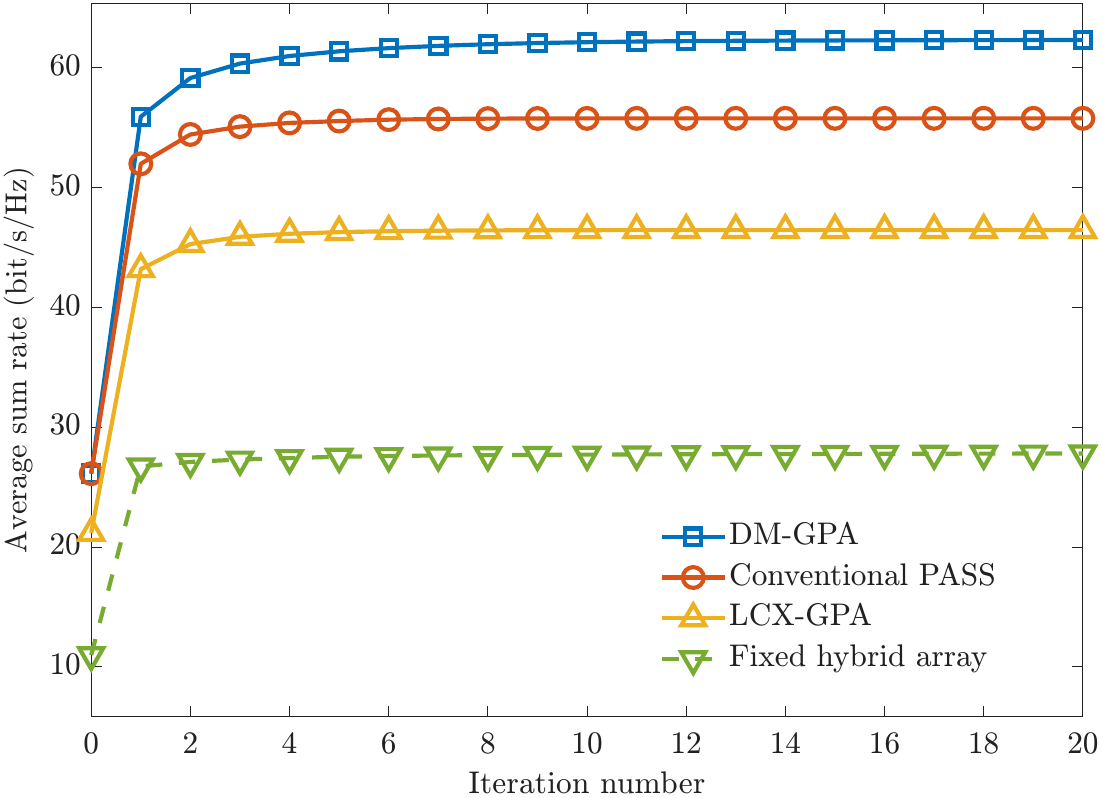}
\caption{Convergence of the proposed {Algorithm~\ref{alg:joint_wmmse}}, where $M=K=8$, $N=96$, $Q=135$, $D_x=10$ m, $h=1$ m, and $\Delta_x=10\lambda_{\rm g}$.}
\label{fig:multiuser_ao_convergence}
\end{figure}

Fig.~\ref{fig:multiuser_ao_convergence} plots the average WSR versus the number of outer AO iterations for DM-GPA and the benchmark schemes. As shown in the figure, the WSR increases monotonically for all schemes, and most of the improvement is achieved within the first few iterations. The DM-GPA curve then gradually stabilizes, consistent with the convergence analysis in Section~\ref{subsec:convergence}. The results further show that DM-GPA achieves the highest WSR after convergence. Although DM-GPA and conventional PASS start from the same initial WSR, DM-GPA attains a larger improvement during the iterations. This result demonstrates the benefit of radiation coefficient optimization when the locations of the activated elements and the digital precoder are jointly optimized.

\subsubsection{Weighted Sum Rate Versus Candidate Spacing}
\label{subsec:numerical_wsr_candidate_spacing}

\begin{figure}[t]
\centering
\includegraphics[width=0.7\columnwidth]
{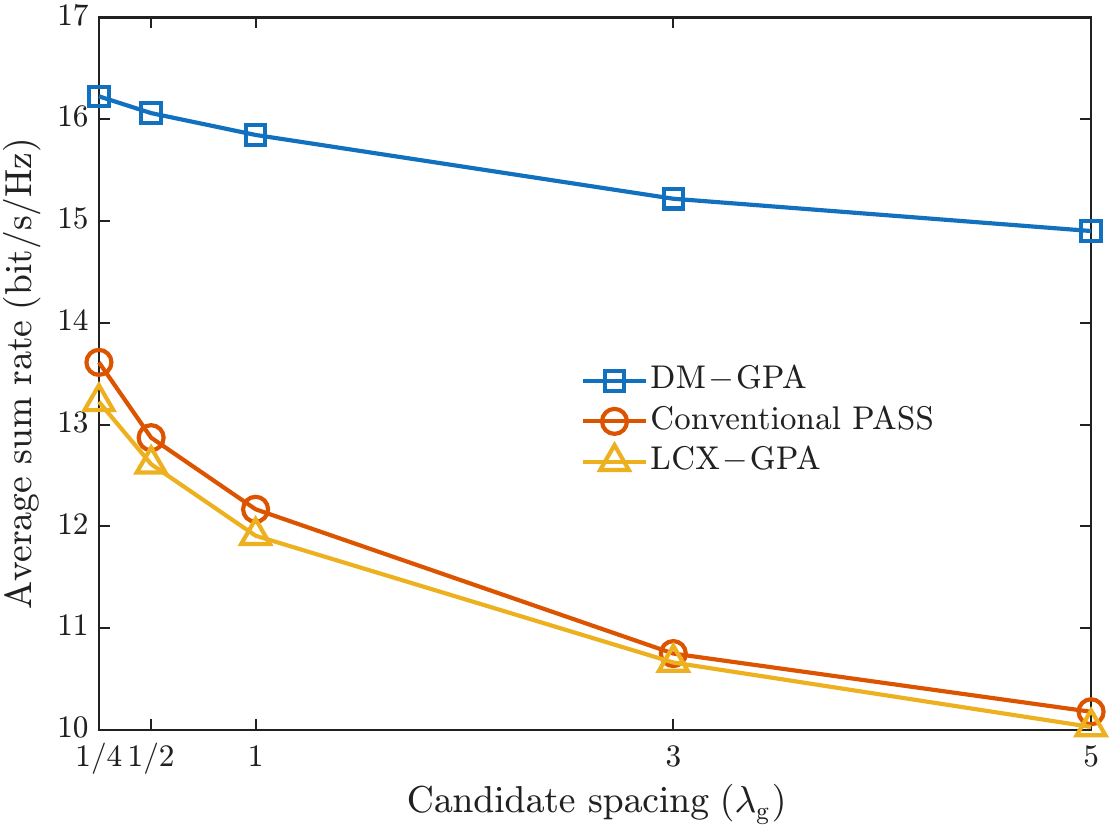}
\caption{Average weighted sum rate versus the candidate-position spacing $\Delta_x$, where $M=K=2$, $N=16$, and $D_x=5$ m. The spacing is normalized by the guided wavelength $\lambda_{\rm g}$.}
\label{fig:wsr_candidate_spacing}
\end{figure}

Fig.~\ref{fig:wsr_candidate_spacing} plots the average WSR versus the spacing $\Delta_x$ between adjacent candidate activation locations. Increasing $\Delta_x$ produces a coarser candidate grid and reduces the number of locations available for pinching beamforming. Consequently, the WSRs of all three architectures decrease as $\Delta_x$ increases. The degradation is substantially smaller for DM-GPA than for conventional PASS and LCX-GPA. For conventional PASS and LCX-GPA, a coarser candidate grid limits the ability of antenna placement to enhance desired signals and suppress inter-user interference. DM-GPA can additionally adjust the radiation amplitudes and phases of the activated elements, which partially compensates for the loss of geometric flexibility on a coarse grid. As a result, its WSR advantage becomes more pronounced at larger candidate spacings. This result confirms that radiation coefficient optimization is particularly beneficial when the available activation locations are sparse.

\subsubsection{Weighted Sum Rate Versus Transmit Power}
\label{subsec:numerical_wsr_power}

\begin{figure}[t]
\centering
\includegraphics[width=0.7\columnwidth]
{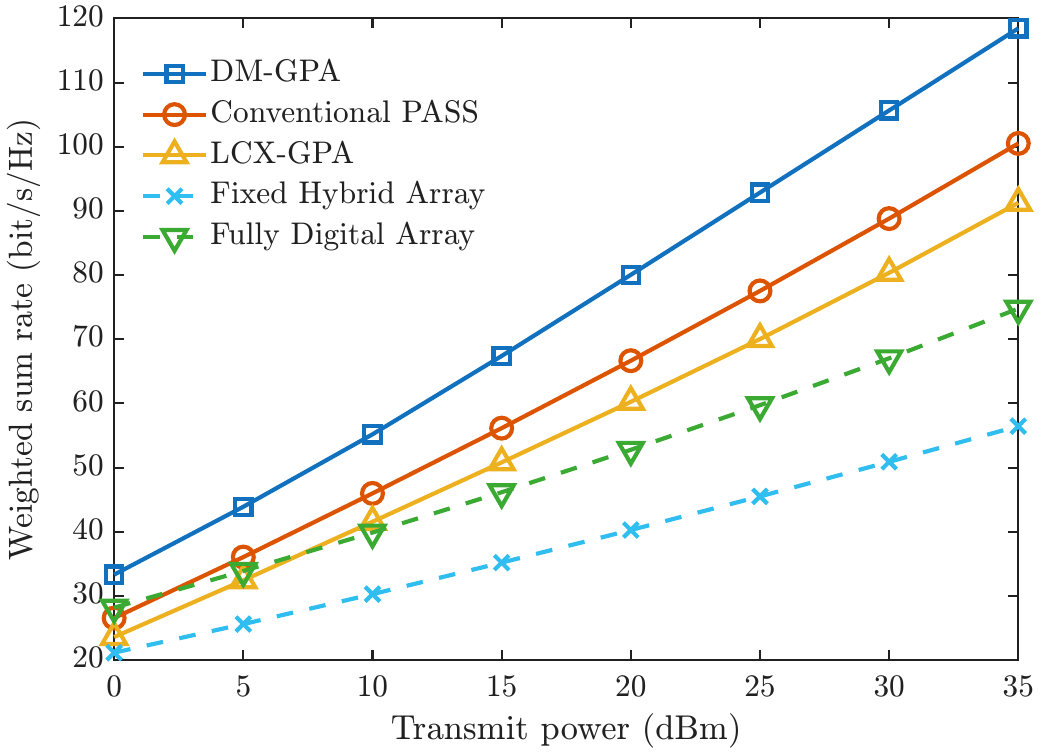}
\caption{Average weighted sum rate versus transmit power, where $M=K=8$, $N=8$, $Q=269$, $D_x=20$ m, and $h=1$ m.}
\label{fig:wsr_transmit_power}
\end{figure}

Fig.~\ref{fig:wsr_transmit_power} plots the average WSR versus the transmit power in the general multiuser scenario. As shown in the figure, the WSR increases with transmit power for all schemes. DM-GPA consistently achieves the highest WSR, and its advantage over the benchmark schemes becomes more pronounced as the transmit power increases. This is because radiation coefficient adjustment provides additional flexibility to suppress inter-user interference, which has a greater impact on the sum rate at high transmit power. Compared with conventional PASS, this improvement is achieved with the same number of RF chains.

It is also observed that the fully digital array achieves a WSR comparable to that of conventional PASS at low transmit power. Its additional digital beamforming degrees of freedom help strengthen the desired signals in this regime. However, conventional PASS and DM-GPA achieve faster WSR growth as the transmit power increases. {The compact fixed array limits the spatial separation of user channels.} In contrast, PASS and DM-GPA can adapt their radiation locations to the user distribution to enhance channel gains and reduce inter-user channel correlation. As a result, their WSR advantage over the fully digital array becomes more pronounced at high transmit power.

\subsubsection{Weighted Sum Rate Versus Number of Users}
\label{subsec:numerical_wsr_users}

\begin{figure}[t]
\centering
\includegraphics[width=0.7\columnwidth]
{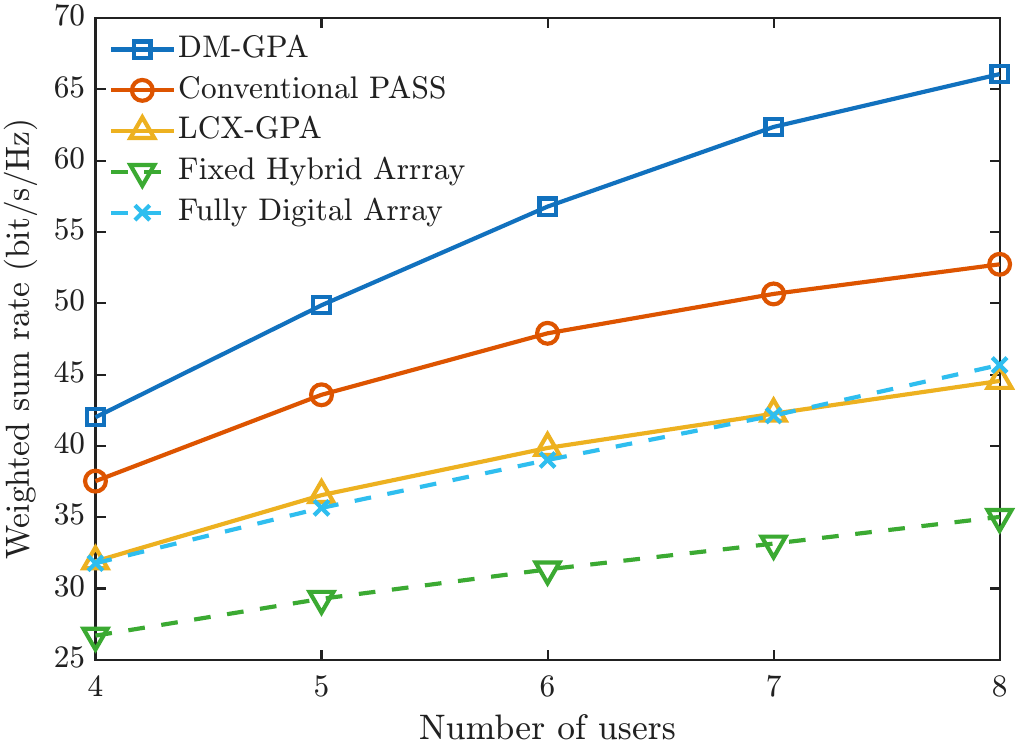}
\caption{Average weighted sum rate versus the number of users $K$, where $M=8$, $N=8$, $Q=269$, $P_{\rm t}=15$ dBm, $D_x=20$ m, and $h=1$ m.}
\label{fig:wsr_number_users}
\end{figure}

Fig.~\ref{fig:wsr_number_users} plots the average WSR versus the number of users $K$. As shown in the figure, the WSR increases with $K$ for all schemes, and DM-GPA consistently achieves the highest WSR. In particular, its advantage over the benchmark schemes becomes more pronounced as more users are served. With a fixed number of RF chains, serving more users makes inter-user interference suppression more challenging. Conventional PASS relies on location selection and digital precoding to manage this interference. In contrast, DM-GPA can additionally adjust the radiation phases of the activated elements through their tunable radiation coefficients to reshape the effective user channels and suppress inter-user interference. As a result, DM-GPA achieves a larger WSR improvement at higher user loads with the same number of RF chains as conventional PASS.

\subsubsection{Weighted Sum Rate Versus Number of Activated Elements}
\label{subsec:numerical_wsr_radiators}

\begin{figure}[t]
\centering
\includegraphics[width=0.7\columnwidth]
{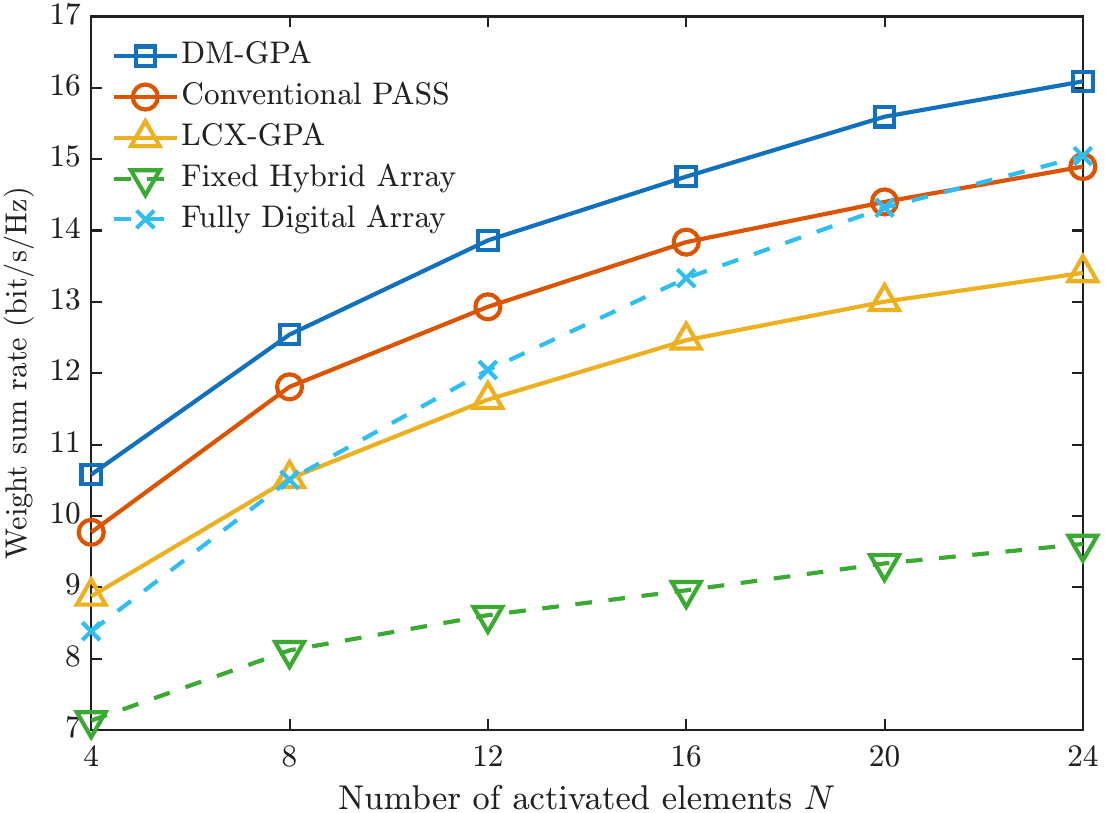}
\caption{Average weighted sum rate versus the number $N$ of activated elements per waveguide, where $M=K=4$ and $D_x=30$ m.}
\label{fig:wsr_selected_radiators}
\end{figure}

Fig.~\ref{fig:wsr_selected_radiators} plots the average WSR versus the number $N$ of activated elements per waveguide. The WSR of DM-GPA improves as more elements are activated and remains higher than that of conventional PASS. This improvement can be explained by two complementary effects. First, more activated elements allow favorable radiation locations near different users to be utilized simultaneously. The additional signal contributions can strengthen the desired signals. Second, each additional activated element introduces a tunable radiation coefficient and thus increases the flexibility to adjust the effective multiuser channels. Through radiation coefficient adjustment, these contributions can be coordinated in amplitude and phase to strengthen desired signals while suppressing inter-user interference. Conventional PASS also benefits from increasing $N$, but its fixed radiation coefficients restrict such coordination. These results demonstrate that DM-GPA can make more effective use of additional activated elements by jointly optimizing their activation locations and radiation coefficients.

\subsubsection{Weighted Sum Rate Versus Waveguide Length}
\label{subsec:numerical_wsr_length}

\begin{figure}[t]
\centering
\includegraphics[width=0.7\columnwidth]
{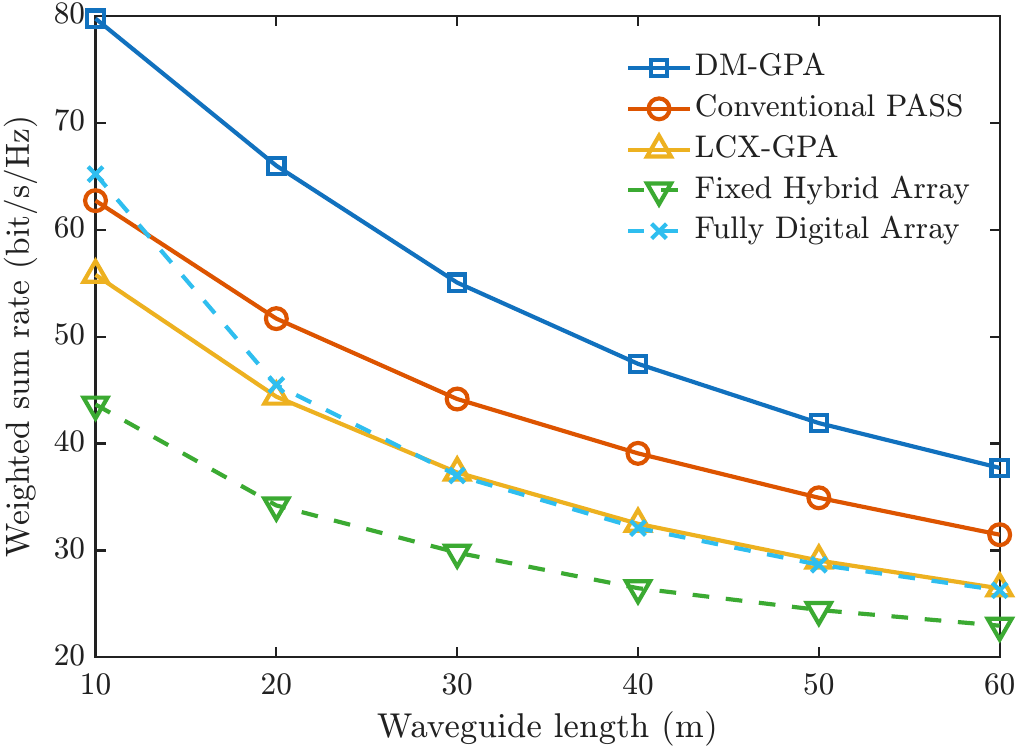}
\caption{Average weighted sum rate versus the waveguide length $D_x$, where $M=K=8$, $N=8$, $P_{\rm t}=15$ dBm, and $h=1$ m.}
\label{fig:wsr_waveguide_length}
\end{figure}

Fig.~\ref{fig:wsr_waveguide_length} plots the average WSR versus the waveguide length $D_x$. As shown in the figure, the WSR decreases for all schemes as $D_x$ increases and the serving region expands. For the fixed arrays, a larger region increases the average user-to-array distance and thus the free-space path loss. In contrast, PASS and DM-GPA can activate elements close to users to shorten the free-space links. However, serving users farther along the waveguides requires longer guided propagation paths and incurs greater in-waveguide attenuation. Despite these losses, DM-GPA consistently outperforms the benchmark schemes through joint location selection and radiation coefficient adjustment.

The comparison with conventional PASS further shows that the WSR improvement of DM-GPA is more pronounced for smaller serving regions. With the number of users fixed, a smaller region results in a higher user density. The reduced spatial separation can increase user channel correlation and make inter-user interference more difficult to suppress. In this case, radiation coefficient adjustment gives DM-GPA additional phase control to strengthen desired signals and suppress inter-user interference. As a result, DM-GPA achieves a larger WSR advantage over conventional PASS in dense user deployments.

\section{Conclusion}
\label{sec:conclusion}

In this article, we proposed a DM-GPA-enabled multiuser downlink framework and investigated joint pinching beamforming and digital precoder design for WSR maximization. For two-user transmission, we derived a closed-form channel orthogonalization design. A lower bound on the sum-rate improvement quantified the tradeoff between interference suppression and channel power loss caused by the Lorentzian constraint. For rank-deficient conventional PASS channels, orthogonalization restored spatial multiplexing and ensured a strictly higher sum rate at sufficiently high transmit powers. The transmit diversity analysis further quantified the gain from coherent combining within each waveguide. For general multiuser transmission, we developed a WMMSE-based AO algorithm. Power normalization yielded a closed-form digital precoder update, while a sequential element-wise search optimized the locations of the activated elements with closed-form radiation coefficient solutions. Numerical results demonstrated that DM-GPA achieved higher WSR than the benchmark schemes and maintained this advantage as the service region expanded. Its WSR increased with more activated elements per waveguide, and its advantage over the PASS benchmarks became more pronounced as the number of candidate locations decreased.
\vspace{-10pt}
\appendices
\section{Proof of Lemma~\ref{prop:exact_channel_diagonalization}}
\label{app:analog_cancellation}

Every nonzero Lorentzian-constrained radiation coefficient in \eqref{eq:lorentzian_theta} can be parameterized as follows:
\begin{equation}
 q_{m,n}=\sin\phi_{m,n}\,e^{\mathrm{j}\phi_{m,n}},
 \qquad 0<\phi_{m,n}<\pi,
\label{eq:lorentzian_phi_parameterization}
\end{equation}
where $\theta_{m,n}=2\phi_{m,n}-\pi/2$ modulo $2\pi$. Substituting this expression into the required radiation coefficient ratio in \eqref{eq:waveguide_null_ratio} gives
\begin{equation}
 \frac{\sin\phi_{m,2}}{\sin\phi_{m,1}}
 e^{\mathrm{j}(\phi_{m,2}-\phi_{m,1})}
 =\varrho_m e^{\mathrm{j}\delta_m}.
\end{equation}
Matching the phases and amplitudes yields $\phi_{m,2}=\phi_{m,1}+\delta_m$ and $\sin(\phi_{m,1}+\delta_m)=\varrho_m\sin\phi_{m,1}$. Expanding the sine and using $\sin\delta_m\neq0$, we obtain
\begin{equation}
 \cot\phi_{m,1}
 =\frac{\varrho_m-\cos\delta_m}{\sin\delta_m}.
\end{equation}
Since $\cot\phi$ is strictly decreasing from $+\infty$ to $-\infty$ on $(0,\pi)$, this equation has the unique solution in \eqref{eq:waveguide_phi_one}. Moreover, $\sin(\phi_{m,1}^{\rm D}+\delta_m)=\varrho_m\sin\phi_{m,1}^{\rm D}>0$ and $\phi_{m,1}^{\rm D}+\delta_m\in(-\pi,2\pi)$ imply $\phi_{m,2}^{\rm D}\in(0,\pi)$. Thus, \eqref{eq:waveguide_closed_form_state} gives the unique pair of nonzero Lorentzian-constrained radiation coefficients satisfying \eqref{eq:waveguide_null_ratio}, which ensures $\widetilde h_{2,1}=\widetilde h_{1,2}=0$.

For the desired link, substituting $q_{m,2}^{\rm D}=-c_{k,m,1}q_{m,1}^{\rm D}/c_{k,m,2}$, with $k\neq m$, gives
\begin{equation}
\begin{aligned}
 \widetilde h_{m,m}^{\rm D}
 &=c_{m,m,1}q_{m,1}^{\rm D}+c_{m,m,2}q_{m,2}^{\rm D}\\
 &=\frac{q_{m,1}^{\rm D}}{c_{k,m,2}}
 \left(c_{m,m,1}c_{k,m,2}-c_{m,m,2}c_{k,m,1}\right)\\
 &=\frac{q_{m,1}^{\rm D} D_m}{c_{k,m,2}}.
\end{aligned}
\end{equation}
The magnitude of the radiation coefficient follows from the cotangent expression as follows:
\begin{equation}
\begin{aligned}
 |q_{m,1}^{\rm D}|
 &=\frac{1}{\sqrt{1+\cot^2\phi_{m,1}^{\rm D}}}\\
 &=\frac{|\sin\delta_m|}
 {\sqrt{1+\varrho_m^2-2\varrho_m\cos\delta_m}}
 =\kappa_m.
\end{aligned}
\end{equation}
Taking the squared magnitude of $\widetilde h_{m,m}^{\rm D}=q_{m,1}^{\rm D} D_m/c_{k,m,2}$ gives \eqref{eq:retained_desired_gain}. {Since $D_m\neq0$ and $\kappa_m>0$, the effective channel gain is strictly positive.} Applying the derived radiation coefficients to both waveguides therefore yields a diagonal effective channel with two nonzero diagonal entries, which completes the proof of Lemma~\ref{prop:exact_channel_diagonalization}.
\vspace{-3pt}
\section{Proof of Theorem~\ref{prop:two_user_rank_one_gain}}
\label{app:rank_one_gain_proof}

For $P_{\rm t}>0$, define $\mathbf V=\mathbf F\mathbf F^{\rm H}/P_{\rm t}$ for a feasible digital precoder. Then $\mathbf V\succeq\mathbf0$ and $\operatorname{Tr}(\mathbf V)\leq1$. When $\rho_{\rm P}=1$, the determinant in \eqref{eq:two_user_cooperative_bound} reduces to $1+\Gamma\operatorname{Tr}(\widetilde{\mathbf H}_{\rm P}\mathbf V\widetilde{\mathbf H}_{\rm P}^{\rm H})$. Since $\mathbf V\preceq\mathbf I_2$, the trace is at most $\|\widetilde{\mathbf H}_{\rm P}\|_{\rm F}^2=A_1+A_2$. Maximizing over the feasible digital precoders proves \eqref{eq:two_user_rank_one_pass_bound}.
Comparing this bound with \eqref{eq:two_user_analog_sum_rate}, the orthogonal DM-GPA construction strictly exceeds the PASS upper bound if and only if
\begin{equation}
 \left(1+\frac{\Gamma S_1^{\rm D}}{2}\right)
 \left(1+\frac{\Gamma S_2^{\rm D}}{2}\right)>1+\Gamma(A_1+A_2).
\end{equation}
Expanding the product and dividing by $\Gamma/4>0$ yields
\begin{equation}
 \Gamma S_1^{\rm D}S_2^{\rm D}>4(A_1+A_2)-2(S_1^{\rm D}+S_2^{\rm D}).
\end{equation}
Since $S_1^{\rm D}S_2^{\rm D}>0$, the left-hand side is positive for every $\Gamma>0$ and grows without bound. If the right-hand side is positive, the inequality holds above a finite positive threshold. If it is nonpositive, it holds for every $\Gamma>0$. Substituting $\Gamma=P_{\rm t}/\sigma^2$ gives \eqref{eq:two_user_rank_one_power_threshold}. At zero transmit power both rates are zero. Finally, the two factors in the DM-GPA rate yield $R_{\rm D}^{\perp}=2\log_2\Gamma+\mathcal O(1)$, whereas \eqref{eq:two_user_rank_one_pass_bound} is $\log_2\Gamma+\mathcal O(1)$. {This establishes the spatial multiplexing and strict rate advantage stated in Theorem~\ref{prop:two_user_rank_one_gain}.}
\vspace{-10pt}
\section{Proof of Theorem~\ref{prop:two_user_rate_bound}}
\label{app:rate_bound_proof}
\vspace{-5pt}
Using the normalized transmit covariance $\mathbf V$ defined in Appendix~\ref{app:rank_one_gain_proof}, the determinant in \eqref{eq:two_user_cooperative_bound} expands as follows:
\begin{equation}
\begin{aligned}
 &\det\left(\mathbf I_2+\Gamma\widetilde{\mathbf H}_{\rm P}\mathbf V\widetilde{\mathbf H}_{\rm P}^{\rm H}\right)\\
 &=1+\Gamma\operatorname{Tr}(\widetilde{\mathbf H}_{\rm P}\mathbf V\widetilde{\mathbf H}_{\rm P}^{\rm H})
 +\Gamma^2|\det\widetilde{\mathbf H}_{\rm P}|^2\det\mathbf V.
\end{aligned}
\end{equation}
As shown in Appendix~\ref{app:rank_one_gain_proof}, the trace term is at most $A_1+A_2$. The eigenvalues of $\mathbf V$ are nonnegative, so their product satisfies $\det\mathbf V\leq[\operatorname{Tr}(\mathbf V)/2]^2\leq1/4$. These bounds hold for every feasible digital precoder. Substituting $|\det\widetilde{\mathbf H}_{\rm P}|^2=A_1A_2(1-\rho_{\rm P}^2)$ proves \eqref{eq:two_user_pass_rate_upper_bound}. Subtracting this upper bound from the achievable DM-GPA rate in \eqref{eq:two_user_analog_sum_rate} gives
\begin{equation}
 R_{\rm D}^{\star}(\mathbf X)-R_{\rm P}^{\star}(\mathbf X)
 \geq R_{\rm D}^{\perp}-U_{\rm P}(\mathbf X)
 =\mathcal L(\Gamma,\mathbf X),
\end{equation}
where combining the two logarithms yields \eqref{eq:two_user_rate_gain_lower_bound}. The feasible-set inclusion of conventional PASS within DM-GPA also holds at each fixed $\mathbf X$ and therefore guarantees a nonnegative optimized rate difference. Combining these two bounds proves \eqref{eq:two_user_fixed_location_rate_gap}. At $P_{\rm t}=0$, both rates and $\mathcal L$ are zero, so the result remains valid.

\section{Proof of Lemma~\ref{cor:two_user_correlation_gain}}
\label{app:correlation_gain_proof}

For $P_{\rm t}>0$, $\mathcal L>0$ in \eqref{eq:two_user_rate_gain_lower_bound} holds if and only if its numerator exceeds its denominator. After subtracting their common constant term and dividing by $\Gamma/4>0$, the equivalent condition can be written as follows:
\begin{equation}
\begin{aligned}
 &\Gamma\big[S_1^{\rm D}S_2^{\rm D}-|\det\widetilde{\mathbf H}_{\rm P}|^2\big]>4(A_1+A_2)-2(S_1^{\rm D}+S_2^{\rm D}).
\end{aligned}
\label{eq:two_user_finite_power_gain_condition}
\end{equation}
Under the stated condition $S_1^{\rm D}S_2^{\rm D}>|\det\widetilde{\mathbf H}_{\rm P}|^2$, the coefficient of $\Gamma$ is strictly positive. Hence, the inequality holds for all sufficiently large $\Gamma$, regardless of the sign of the right-hand side. Since $\mathcal L=R_{\rm D}^{\perp}-U_{\rm P}$ and $R_{\rm P}^{\star}\leq U_{\rm P}$, this proves the strict rate advantage of the orthogonal construction.

For a fixed channel with $\rho_{\rm P}<1$, dividing the numerator and denominator in \eqref{eq:two_user_rate_gain_lower_bound} by $\Gamma^2/4$ gives
\begin{equation}
 \mathcal L=\log_2\!\left[
 \frac{S_1^{\rm D}S_2^{\rm D}+2\Gamma^{-1}(S_1^{\rm D}+S_2^{\rm D})+4\Gamma^{-2}}
 {A_1A_2(1-\rho_{\rm P}^2)+4\Gamma^{-1}(A_1+A_2)+4\Gamma^{-2}}
 \right].
\end{equation}
Taking $\Gamma\rightarrow\infty$ yields \eqref{eq:two_user_asymptotic_rate_gain}, which is positive under the assumed gain condition.

Finally, hold $A_1,A_2,S_1^{\rm D},S_2^{\rm D}$ and $\Gamma$ fixed. The numerator in \eqref{eq:two_user_rate_gain_lower_bound} is independent of $\rho_{\rm P}$, while the derivative of its denominator is $-\Gamma^2A_1A_2\rho_{\rm P}/2$. Applying the derivative of $-\log_2(\cdot)$ gives \eqref{eq:two_user_correlation_gain_derivative}, which is strictly positive for $0<\rho_{\rm P}<1$. This completes the proof of Lemma~\ref{cor:two_user_correlation_gain}.

\vspace{-5pt}
\section{Proof of Lemma~\ref{lemma:wmmse_equivalence}}
\label{app:wmmse_proof}

For fixed $(\mathbf F,\mathbf X,\mathbf Z)$, completing the square in \eqref{eq:mse} using $T_k$ from \eqref{eq:total_received_power} gives
\begin{align}
 e_k
 &=T_k|u_k|^2-2\operatorname{Re}\{u_k^*\xi_{k,k}\}+1\notag\\
 &=T_k\left|u_k-\frac{\xi_{k,k}}{T_k}\right|^2
 +1-\frac{|\xi_{k,k}|^2}{T_k}.
\label{eq:mse_complete_square}
\end{align}
The first term in \eqref{eq:mse_complete_square} is nonnegative and vanishes at $u_k^\star=\xi_{k,k}/T_k$. This proves \eqref{eq:mmse_total}. Substitution yields
\begin{align}
 e_k^{\rm MMSE}
 &=1-\frac{|\xi_{k,k}|^2}{T_k}\notag\\
 &=\frac{\sum_{i\neq k}|\xi_{k,i}|^2
 +\dfrac{\sigma_k^2}{P_{\rm t}}\|\mathbf F\|_{\rm F}^2}
 {|\xi_{k,k}|^2+\sum_{i\neq k}|\xi_{k,i}|^2
 +\dfrac{\sigma_k^2}{P_{\rm t}}\|\mathbf F\|_{\rm F}^2}
 =\frac{1}{1+\overline\gamma_k}.
\label{eq:appendix_mmse}
\end{align}

For fixed $e_k>0$, let $\phi_k(v_k)=v_ke_k-\ln v_k$ for $v_k>0$. Differentiation gives
\begin{equation}
 \phi_k'(v_k)=e_k-\frac{1}{v_k},\qquad
 \phi_k''(v_k)=\frac{1}{v_k^2}>0.
\label{eq:weight_derivatives}
\end{equation}
Thus, $\phi_k$ is strictly convex and has the unique minimizer $v_k^\star=1/e_k$. With $e_k=e_k^{\rm MMSE}$,
\begin{equation}
 v_k^\star=\frac{1}{e_k^{\rm MMSE}}=1+\overline\gamma_k,
\end{equation}
and
\begin{align}
 \min_{v_k>0}\phi_k(v_k)
 &=1+\ln e_k^{\rm MMSE}=1-\ln(1+\overline\gamma_k).
\label{eq:appendix_weight_minimum}
\end{align}
Consequently,
\begin{align}
 \min_{\mathbf u,\mathbf v\succ\mathbf 0}\mathcal{J}
 &=\sum_{k=1}^{K}\omega_k[1-\ln(1+\overline\gamma_k)]\notag\\
 &=\sum_{k=1}^{K}\omega_k
 -\ln(2)\sum_{k=1}^{K}\omega_k\log_2(1+\overline\gamma_k),
\end{align}
which proves Lemma~\ref{lemma:wmmse_equivalence}.

\vspace{-5pt}
\bibliographystyle{IEEEtran}
\bibliography{reference}

\end{document}